\documentclass[final,5p,times,twocolumn]{elsarticle}

\usepackage{graphics}
\usepackage{graphicx}
\usepackage{amssymb}
\usepackage[dvipsnames]{xcolor}
\usepackage[colorlinks=true,urlcolor=Blue,linkcolor=Blue]{hyperref}
\usepackage[all]{hypcap}
\def\Journal#1#2#3#4{{#1} {#2} (#4) #3 }

\def\NPA{{\em Nucl. Phys.} A}

\def\PLB{{\em Phys. Lett.} B}

\def\PRL{\em Phys. Rev. Lett.}

\def\PREP{\em Phys. Rep.}
\def\PRA{{\em Phys. Rev.} A}

\def\PRC{{\em Phys. Rev.} C}

\def\ZPA{{\em Z. Phys.} A}

\def\RMP{{\em Rev. Mod. Phys.}}

\def\NIMA{{\em Nucl. Instr. Meth. Phys. Research} A}
\def\NIMB{{\em Nucl. Instr. Meth. Phys. Research} B}
\def\jpg{{\em J. Phys.} G}
\def\PS{{\em Phys. Scripta}}

\def\RPP{{\em Rep. Prog. Phys.}}
\def\HI{{\em Hyperfine Interact.}}
\def\EPJA{{\em Eur. Phys. J.} A}
\def\ADNDT{{\em At. Data Nucl. Data Tables}}
\def\APPB{{\em Acta Phys. Polonica} B}

\journal{Nuclear Instruments and Methods in Physics Research, Section B}

\begin{document}

\begin{frontmatter}



\title{Nuclear Physics Experiments with Ion Storage Rings}


\author[GSI,UNIHD]{Yu.~A.~Litvinov} %\fnref{fn1}}%
\author[TUM]{S.~Bishop}%
\author[MPI]{K.~Blaum}%
\author[GSI]{F.~Bosch}%
\author[EMMI]{C.~Brandau}%
%\author[ENG]{Peter~Butler}%
\author[TEXAS]{L.~X.~Chen}%
\author[GSI,UniG]{I.~Dillmann}%
\author[GSI]{P.~Egelhof}%
\author[GSI,UniG]{H.~Geissel}%
\author[GSI,UniF]{R.~E.~Grisenti}%
\author[GSI,UniF]{S.~Hagmann}%
\author[GSI]{M.~Heil}%
\author[SWE]{A.~Heinz}%
\author[KVI]{N.~Kalantar-Nayestanaki}%%
\author[GSI,UniG]{R.~Kn{\"o}bel}%
\author[GSI]{C.~Kozhuharov}%
\author[GSI]{M.~Lestinsky}%
\author[IMP]{X.~W.~Ma}%
\author[SWE]{T.~Nilsson}%
\author[GSI]{F.~Nolden}%
\author[label3]{A.~Ozawa}
\author[TSR]{R.~Raabe}%
\author[ANU]{M.~W.~Reed}%
\author[UniF]{R.~Reifarth}%
\author[GSI,UniF]{M.~S.~Sanjari}%
\author[Cal]{D.~Schneider}%
\author[GSI]{H.~Simon}%
\author[GSI]{M.~Steck}%
\author[GSI, HIJ, UniJ]{T.~St{\"o}hlker}%
\author[Bei]{B.~H.~Sun}%
\author[IMP]{X.~L.~Tu}%
\author[label2]{T.~Uesaka}
\author[UniS]{P.~M.~Walker}%
\author[label2]{M.~Wakasugi}
\author[GSI]{H.~Weick}%
\author[GSI,MPI]{N.~Winckler}%
\author[UniE]{P.~J.~Woods}%
\author[IMP]{H.~S.~Xu}%
\author[label1]{T.~Yamaguchi}
\author[label2]{Y.~Yamaguchi}
\author[IMP]{Y.~H.~Zhang}%

\address[GSI]{GSI Helmholtzzentrum f\"ur Schwerionenforschung (GSI), 64291 Darmstadt, Germany}%
\address[UNIHD]{Ruprecht-Karls Universit{\"a}t Heidelberg, 69120 Heidelberg, Germany}%
\address[TUM]{Technische Universit{\"a}t M{\"u}nchen, 85748 Garching, Germany}%
\address[MPI]{Max-Planck-Institut f\"ur Kernphysik (MPIK) , 69117 Heidelberg, Germany}%
\address[EMMI]{ExtreMe Matter Institute EMMI, 64291 Darmstadt, Germany}%
\address[TEXAS]{Cyclotron Institute, Texas A\&M University, College Station, Texas 77843, USA}
\address[UniG]{Justus-Liebig Universit{\"a}t, 35392 Gie{\ss}en, Germany}%
\address[UniF]{J.~W.-Goethe Universit{\"a}t, 60438 Frankfurt, Germany}%
\address[SWE]{Chalmers University of Technology, SE-412 96 Gothenburg, Sweden}%
\address[KVI]{Kernfysisch Versneller Institute (KVI), University of Groningen, 9747 AA Groningen, The Netherlands}%
\address[IMP]{Institute of Modern Physics, Chinese Academy of Sciences (IMP), Lanzhou 730000, China}%
%\address[Cha]{...}%
\address[label3]{Institute of Physics, University of Tsukuba, Ibaraki 305-8571, Japan}%
\address[TSR]{Instituut voor Kern- en Stralingsfysica, KU Leuven, 3001 Leuven, Belgium}%
\address[ANU]{Australian National University, Canberra ACT 0200, Australia}%
\address[Cal]{Lawrence Livermore National Laboratory, Livermore, CA 94551, USA}%
\address[HIJ]{Helmholtz-Institut Jena, 07743 Jena, Germany}%
\address[UniJ]{Friedrich-Schiller-Universit{\"a}t Jena, 07737 Jena, Germany}%
\address[Bei]{School of Physics \& Nucl. Energy Engineering, Beihang Univ., 100191 Beijing, China}%
\address[label2]{RIKEN Nishina Center, RIKEN, Wako, Saitama 351-0198, Japan}%
\address[UniS]{Department of Physics, University of Surrey, Guildford, GU2 7XH, UK}%
\address[UniE]{School of Physics \& Astronomy, The University of Edinburgh, Edinburgh EH9 3JZ, UK}%
\address[label1]{Department of Physics, Saitama University, Saitama 338-8570, Japan}%

%\fntext[fn1]{Electronic address: Y.Litvinov@GSI.de}%

\begin{abstract}
In the last two decades a number of nuclear structure and astrophysics experiments were performed at heavy-ion storage rings
employing unique experimental conditions offered by such machines. 
Furthermore, building on the experience gained at the two facilities presently in operation, several new storage ring projects were launched worldwide.
This contribution is intended to provide a brief review of the fast growing field of nuclear structure and astrophysics research at storage rings.
\end{abstract}


\end{frontmatter}

% \linenumbers

\section{Introduction}
\label{intro}
%{\bf Markus, Fritz, Robert et al., please check critically}\\

Ion storage rings offer unique experimental conditions for precision experiments 
with stable and--if coupled to radioactive beam facilities--also with exotic nuclei.
The research potential is enormous which was demonstrated in the last years by a number of successful experiments.
Presently, there are two such machines in operation, namely 
the experimental storage ring ESR~\cite{ESR} at GSI~\cite{GSI} and
%the GSI Helmholtz Centre in Darmstadt and 
the cooler-storage ring CSRe~\cite{CSRe} at IMP~\cite{IMP}.
%at the Institute of Modern Physics, Chinese Academy of Sciences in Lanzhou.
Several new storage ring facilities are being constructed or are in a planning phase.

%%%%%%%%%%%%%%%%%%%%%%%%%%%%%%%%%%%%%%%%%%%%%%%%%%
\begin{figure*}[t!]
\centering\includegraphics[angle=-0,width=0.8\textwidth]{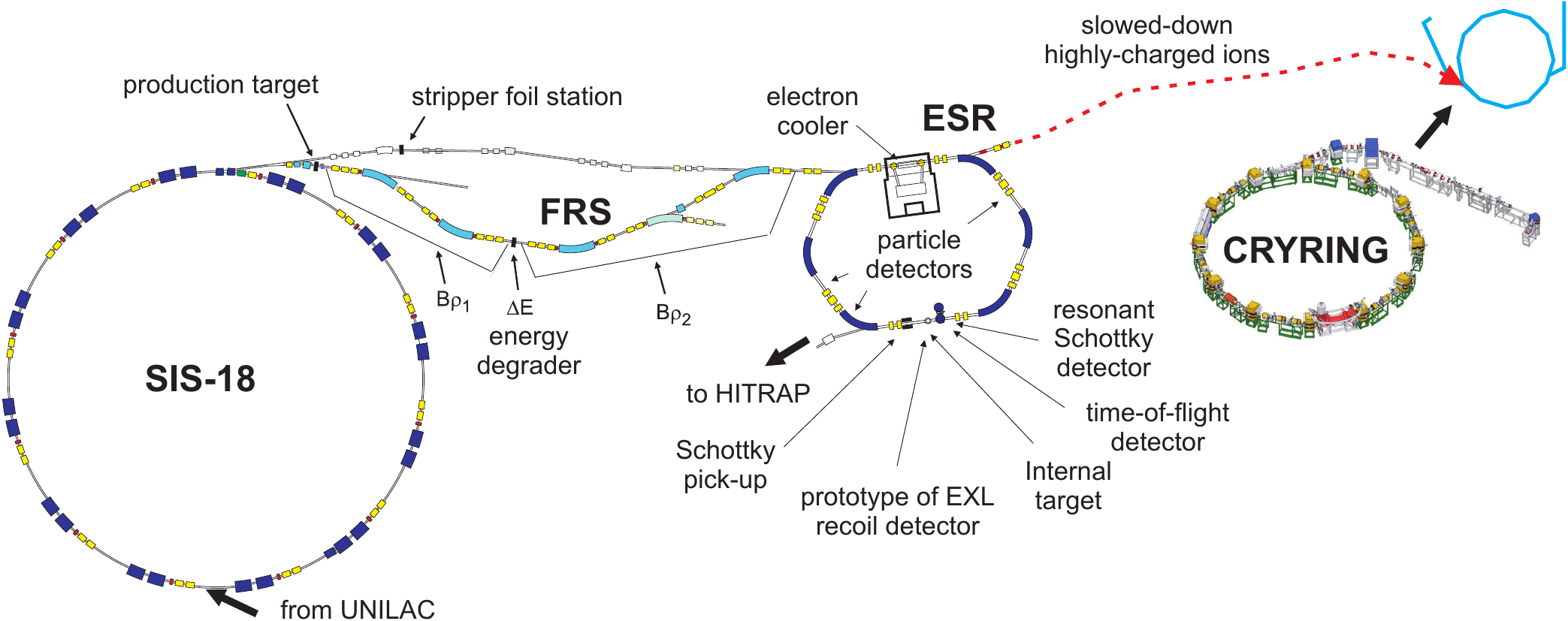}
\caption{(Colour online) The FRS-ESR arrangement at GSI. The two stages for in-flight $B\rho-\Delta{\rm E}-B\rho$ 
separation and the energy degrader in the FRS~\cite{FRS} as well  as 
the electron cooler and the main detection systems in the ESR are shown. 
The preferred location of the low-energy storage ring CRYRING (see insert) is indicated together with the beam-line connecting it to the ESR for transporting
decelerated highly-charged stable or radioactive ions. Adopted from~\cite{Phil100, BoLiSt}}
\label{frs_esr}
\end{figure*}
%%%%%%%%%%%%%%%%%%%%%%%%%%%%%%%%%%%%%%%%%%%%%%%%%%
Owing to the ultra-high vacuum in the order of $10^{-10}-10^{-12}$~mbar, exotic nuclides 
can be stored for extended periods of time reaching--dependent on the nuclear half-lives, 
atomic charge states and kinetic energy--several hours~\cite{LiBo}.

The most important capability of storage rings is beam cooling, which allows the reduction of the energy spread of the stored
ions induced, for instance, by the production reaction process, the interaction with internal targets, or due to recoiling in a decay~\cite{No2}.
Stochastic~\cite{No-NIMA} and electron~\cite{St-NIMA} cooling techniques 
are applied which enable beams with the highest phase-space density. % and well-defined velocity.
For electron cooled beams, the momentum spread of stored ions $\delta p/p$ is determined by the equilibrium between 
the Coulomb interactions with cold electrons in the cooler device and the intra-beam scattering~\cite{Poth}.
The latter increases rapidly with increasing beam intensity and the atomic charge state of the ions.
A momentum spread of $\delta p/p\sim10^{-5}$ is achievable for $10^7$ fully-ionised $^{238}$U$^{92+}$ ions, 
which is reduced by an order of magnitude for a few $10^3$ stored ions.
At even smaller numbers of stored ions a ``phase transition'' occurs, 
at which the intra-beam scattering is ``switched off'' and so-called ``crystalline beams'' with $\delta p/p$ of a few $10^{-7}$ are observed~\cite{SteckS-1, SteckS-2}.
For a large initial velocity spread, the electron cooling can take up to a few minutes.
However, if the stochastic pre-cooling, which is capable
of reducing the momentum spread quickly to
 %to quickly reduce the momentum spread to 
 $\delta p/p\sim10^{-4}$, 
is applied prior to the electron cooling, then the overall cooling time can be reduced to a few seconds~\cite{Geissel04}.
Thus, cooled beams of radioactive ions with half-lives in the order of one second or longer can be prepared.
Recently, laser cooling~\cite{LaserCooling}, pioneered in the test storage ring TSR in Heidelberg~\cite{laser}, 
was successfully demonstrated in the ESR with stored relativistic carbon ions~\cite{laser-ESR,LC2012,add1}.

For in-ring reaction experiments (see Section~\ref{S:reactions}), internal targets can be used~\cite{gas_targets}. 
Normally gas or cluster targets are employed~\cite{gjESR, Grisenti}.
A wide range of gases like H$_2$, d, $^3$He, $^4$He, Ar, Xe, CH$_4$, etc. can be utilised. 
The targets have small dimensions of less than 10~mm at the interaction zone and are very thin having thicknesses of $<10^{15}$~atoms/cm$^2$.
This enables a very high angular and energy resolution in reaction measurements
and access to measurements at very low momentum 
transfer where low energy recoil particles need to be detected~\cite{PEPS,FAIRBS}.
Furthermore, such targets are windowless, which is essential since no corrections are needed to subtract the background from the interactions in windows.
Although the targets are thin, relatively high luminosities are obtained owing to high revolution frequencies of the ions in a ring, which are typically $10^5-10^6$~Hz.
Assuming a stored beam with a moderate intensity of $10^5$ ions, a luminosity of $10^{25}-10^{26}$~/cm$^2\cdot$s 
can be achieved.
We note that densities in excess of $10^{14}$~atoms/cm$^2$ can be achieved by using droplet targets, 
which, however, affects the excellent properties of the cooled beams~\cite{Petridis-NIM}.

Heavy-ion storage rings are typically high-acceptance devices which allow for storing many different nuclear species at the same time.
This is successfully used in in-ring mass measurements (see Section~\ref{S:masses}), where by measuring the revolution frequencies of the ions 
of interest and by comparing them to the revolution frequencies of nuclides with well-known masses, the unknown masses can accurately be determined~\cite{FGM}.
Often a single ion is sufficient to obtain its mass, which is essential if measurements of very exotic nuclei with small production rates are considered~\cite{Li132}.

Another application of the large storage acceptance of the rings is the possibility to study radioactive decays~\cite{Li04} (see Section~\ref{S:lifetimes}).
Since the mass of an ion is changed after its decay, the revolution frequency is changed as well.
Therefore, time- and frequency-resolved current measurements in a ring enable precision life-time and branching ratio measurements.
The latter are also possible with single stored ions~\cite{LiBo}.
A unique feature is the possibility to unambiguously monitor the charge state of the stored ions during the entire measurement time.
This is important for decay studies of highy-charged ions, like, 
for instance, fully-ionised, hydrogen- (H-like), helium- (He-like), or lithium-like (Li-like) ions.

Storage rings can be used as synchrotrons to accelerate or decelerate stored beams to a required energy~\cite{SteckEPAC}.
This capability turns out to be essential for nuclear reaction studies with internal targets.
A striking example are the nuclear astrophysics measurements that can be performed directly in the Gamow window of the corresponding astrophysical process.
Stored highly-charged radionuclides can presently be slowed down at the ESR only.
The available beam energies are from about 400~MeV/u--typical injection energy--down to about 4~MeV/u. 
The latter is the extraction energy towards the external HITRAP setup~\cite{HITRAP}.
Stored and cooled exotic beams in the energy range of a few MeV/u will also be available 
once the storage ring project~\cite{TSR} at ISOLDE~\cite{ISOLDE} at CERN, Geneva is realised.
By coupling the CRYRING to the ESR--the CRYRING@ESR project~\cite{CRYRING0}--beams 
of highly-charged ions with energies as low as several tens of keV/u will be available.
The construction of the high-energy storage ring HESR at the future FAIR in Germany and 
the realisation of the HIAF project in China, will enable stored ion beams with energies up to 5~GeV/u or even higher~\cite{SPARC_HESR}.

In this contribution we briefly review nuclear physics experiments 
at the present ESR and CSRe storage rings and
sketch the new storage ring projects being pursued in the world.

\section{Existing Heavy-Ion Storage Ring Facilities}
%{\bf All, please have a look}\\

The only operating ion storage ring facilities are located at GSI and IMP~\cite{YL-ML}.
At GSI, the radioactive ion beam facility is a combination of the high-energy heavy-ion synchrotron SIS~\cite{SIS}, 
the in-flight fragment separator FRS~\cite{FRS}, and the cooler-storage ring ESR~\cite{ESR}. 
The facility is schematically illustrated in Fig.~\ref{frs_esr}. 
Here, the intense beams of any stable isotope from protons up to uranium can be accelerated by 
the SIS to a maximum magnetic rigidity of $B\rho=mv\gamma/q=18$~Tm,
where $B$ is the applied magnetic field, $\rho$, $m/q$ and $v$ are the bending radius, the mass-over-charge ratio 
and the velocity of the accelerated particles, respectively, 
while $\gamma$ is the relativistic Lorentz factor.
Relativistic fragments with energies of several hundreds of 
MeV/u are produced mainly through fragmentation 
of primary beam projectiles in thick production targets. 
In the case of uranium primary beams also projectile 
fission is used for the production of neutron-rich nuclei. 
Typically beryllium targets with thicknesses of $1-8$~g/cm$^2$ are employed. 
Secondary beams are separated in flight in the FRS 
within about 150~ns and are then injected into the ESR. 
The maximal magnetic rigidity of the ESR is 10~Tm.
Dependent on the specific experimental requirements, 
cocktail or clean mono-isotopic beams can be prepared 
with the FRS by employing magnetic rigidity $(B\rho)$ analysis 
and atomic energy loss $(\Delta E)$ in specially shaped solid state degraders. 
Cocktail beams are ideally suited for the in-ring mass measurements~\cite{FGM}, whereas
pure mono-isotopic beams are often needed for lifetime determination or reaction studies.
The only significant disadvantage of the 
FRS-ESR facility is the low injection efficiency into the ESR, which is of the order of a few percents~\cite{Ra-NPA}.

%%%%%%%%%%%%%%%%%%%%%%%%%%%%%%%%%%%%%%%%%%%%%%%%%%
\begin{figure}[t!]
\centering\includegraphics[angle=-0,width=0.4\textwidth]{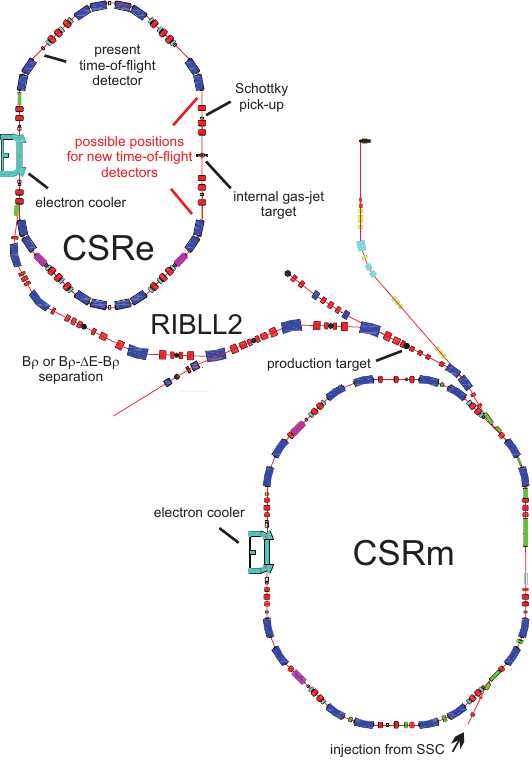}
\caption{(Colour online) Schematic view of the radioactive beam facility at the IMP in Lanzhou~\cite{CSRe, Xiao}. 
Primary beams are accelerated by the synchrotron CSRm, fast extracted and focused on the production target in front of the in-flight fragment separator RIBLL2.
Projectile fragments are separated in the RIBLL2 and injected and stored in the storage ring CSRe. 
Revolution frequencies of the electron-cooled ions can be measured using the Schottky pick-up. 
If the CSRe is tuned into the isochronous ion-optical mode then the revolution frequencies 
can be obtained also for uncooled particles by using dedicated time-of-flight detectors. Adopted from~\cite{CSRe}.}
\label{csre}
\end{figure}
%%%%%%%%%%%%%%%%%%%%%%%%%%%%%%%%%%%%%%%%%%%%%%%%%%

\begin{figure}[t!]
\centering\includegraphics[angle=90,width=0.4\textwidth]{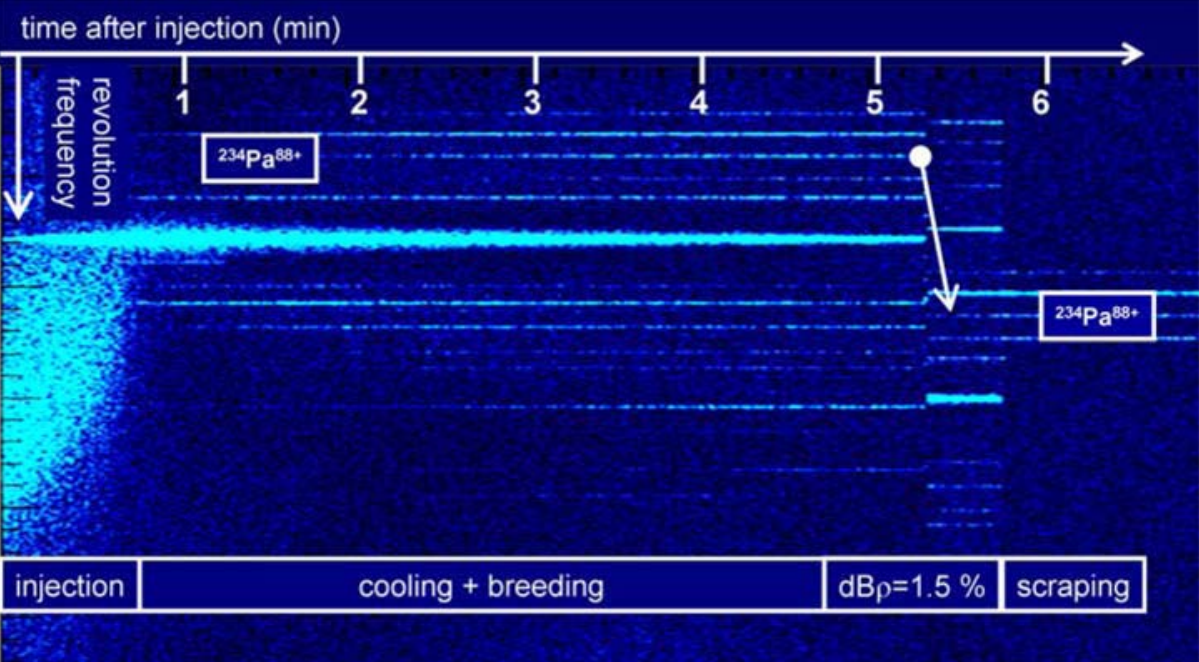}
\caption{%
(Colour online) Time-resolved Schottky frequency spectra illustrating the preparation of a Li-like $^{234}$Pa$^{88+}$ beam in the ESR. 
The y-axis denotes the time after injection and the x-axis the revolution frequency of the stored ions,
which for cooled ions reflects directly the mass-over-charge ratios $m/q$ of the ions. 
%The intensity is plotted in a logarithmic scale.
The radionuclides were produced in a production target inserted in the stripper-foil station 
of the direct transport line.
% which was optimised for the maximal transmission of Li-like $^{237}$U$^{89+}$ ions. 
A long waiting time was necessary in this experiment to breed the Li-like $^{234}$Pa$^{88+}$ beam 
via atomic electron capture on He-like $^{234}$Pa$^{89+}$ ions in the electron cooler.
The arrow indicates the moving of the beam of interest onto the central orbit of the ESR by ramping the ESR magnets.
After the insertion of mechanical scrapers, the isotopes $^{232}$Th$^{87+}$, 
$^{237}$U$^{89+}$, $^{234}$Pa$^{88+}$ and $^{231}$Th$^{87+}$ (left to right) remained in the ring. 
Further cleaning of the Li-like $^{237}$U$^{89+}$ beam, although feasible, 
was not required in this experiment. Adopted from~\cite{{Brandau10}}.
%In close vicinity of the uranium and the protactinium lines, i.e., with almost identical m/q ratio, contaminants from 229Ac86+ and 226Ra85+ are in the ring, 
%but with much lower intensity [44]. One should note that the intensity scale is logarithmic
}%
\label{separation}
\end{figure}

\begin{figure}[t!]
\centering\includegraphics[angle=-0,width=0.4\textwidth]{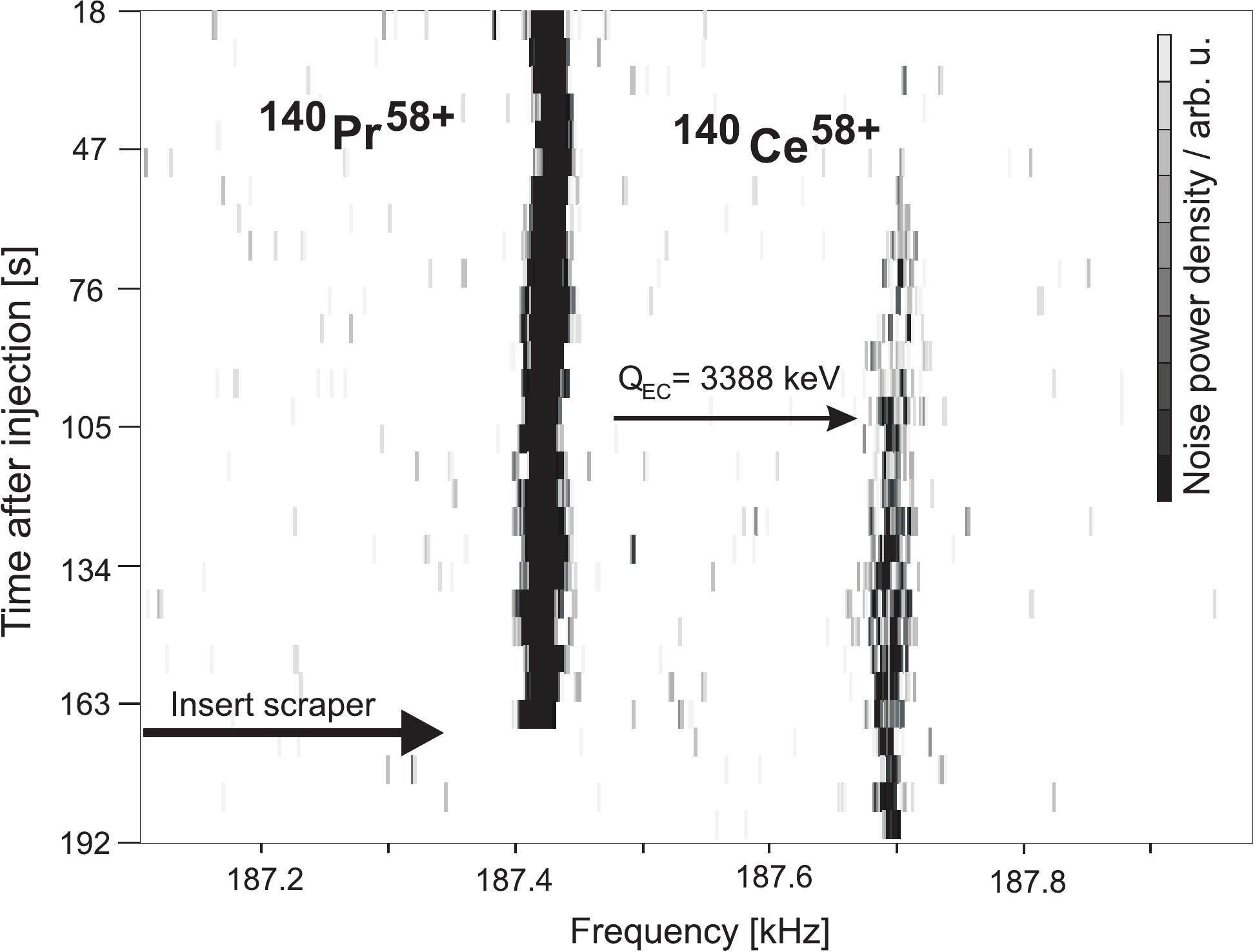}
\caption{%
Time-resolved Schottky frequency spectra of stored and cooled H-like $^{140}$Pr$^{58+}$ ions and bare $^{140}$Ce$^{58+}$ nuclei.
These isobars are separated in energy by 3.4~MeV. 
The heavier isobaric component is removed by inserting a scraper inside the ESR aperture. 
The corresponding time is indicated by the arrow. Adopted from~\cite{Sch-HI173}.
}%
\label{scraping}
\end{figure}

Exotic nuclei can also be produced in the direct transfer line connecting 
SIS and ESR, see Fig.~\ref{frs_esr}, by installing a production target in the stripper-foil station. 
%, which is an important option since the FRS is normally overbooked with experiments not involving ESR. 
%A rough separation of the nuclides of interest can be done with the dipole magnets of the transfer line.
The separation from inevitable contaminants is achieved inside the storage ring~\cite{Brandau10}.
% on the cooled ions beams.
This is illustrated in Fig.~\ref{separation}.
% where the $^{238}$U projectile 
%fragments are mechanically separated in the ESR, 
%such that a beam of Li-like $^{234}$Pa$^{88+}$ ions is stored on the central orbit of the ESR. 
%purified from majority of contaminants.
%%%%%%%%%%%%%%%%%%%%%%%%%%%%%%%%%%%%%%%%%%%%%%%%%%
%%%%%%%%%%%%%%%%%%%%%%%%%%%%%%%%%%%%%%%%%%%%%%%%%%
If required a pure mono-isotopic or--dependent on the excitation energy--even a pure mono-isomeric beam 
can be prepared in the ESR owing to the high resolving power achievable with electron-cooled beams.
This is illustrated in Fig.~\ref{scraping}, where the heavier of two $A = 140$ isobars--separated 
in energy by merely 3.4~MeV--is removed from the ESR~\cite{Sch-HI173}. 

%%%%%%%%%%%%%%%%%%%%%%%%%%%%%%%%%%%%%%%%%%%%%%%%%%
%%%%%%%%%%%%%%%%%%%%%%%%%%%%%%%%%%%%%%%%%%%%%%%%%%
%%%%%%%%%%%%%%%%%%%%%%%%%%%%%%%%%%%%%%%%%%%%%%%%%%%%%%%%%
\begin{figure}[h]
\begin{center}
\includegraphics*[width=0.4\textwidth]{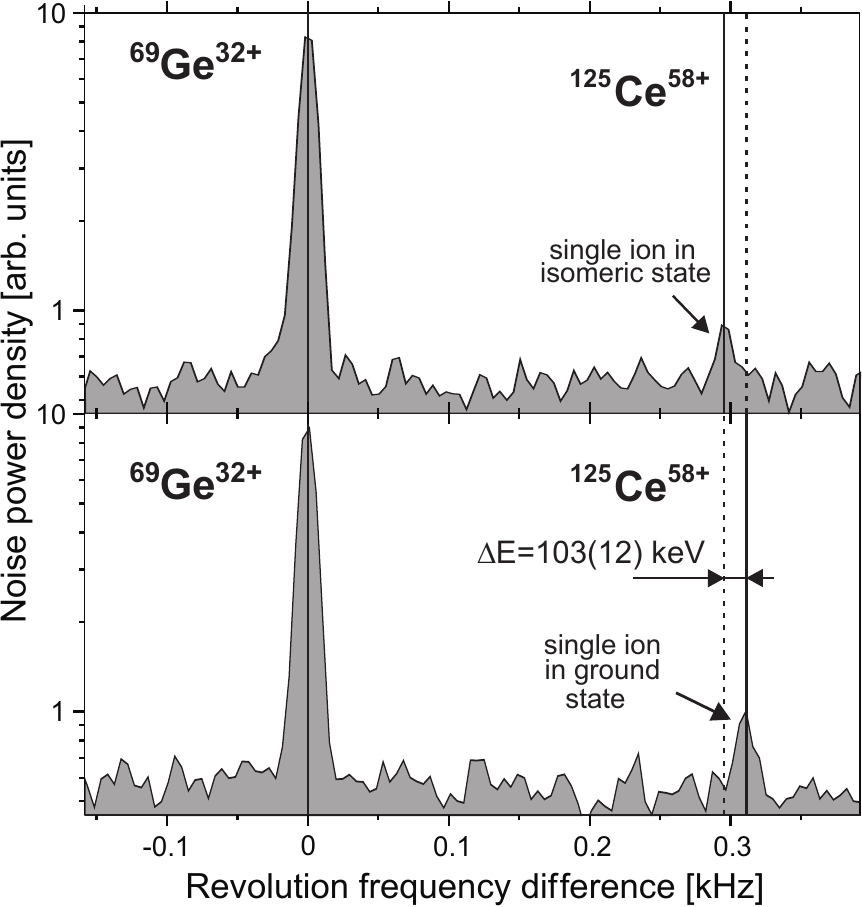}
\end{center}
{\caption{Schottky frequency spectra of two measurements in the ESR 
of single stored fully-ionised $^{125}$Ce$^{58+}$ ions in the isomeric (upper panel) and ground (lower panel) states.
The frequency peak of $^{69}$Ge$^{32+}$ ions can be used as a reference. The frequency difference between the two single $^{125}$Ce ions corresponds
to the isomeric excitation energy of $E^*=103(12)$~keV. Taken from~\cite{Sun07}. } 
\label{125ce}}
\end{figure}
%%%%%%%%%%%%%%%%%%%%%%%%%%%%%%%%%%%%%%%%%%%%%%%%%%%%%%%%%%

A similar scheme is realised at IMP, where the high energy part of the facility consists of the synchrotron CSRm,
which is coupled to the storage ring CSRe via the in-flight separator RIBLL2.
The facility is schematically illustrated in Fig.~\ref{csre}~\cite{Xiao}.
%Some characteristics of the ESR and CSRe rings are summarised in Table~\ref{tab rings}.
As in the case of SIS-FRS-ESR, the primary beams extracted from the CSRm are fragmented in a beryllium production target 
in front of RIBLL2, separated in flight and are injected into the CSRe.
The CSRe has a mean circumference of 128.8~m and a maximal magnetic rigidity of 8.4~Tm. 
An advantageous difference of the CSRe in comparison to the ESR is that it has two longer straight sections 
which allow for more flexibility in designing reaction experiments and provide more space for detector setups. 
Several experiments are being considered in the CSRe which will profit from these properties.

Major parameters of the CSRe and ESR storage rings are listed in Table~\ref{rings}.
%------------
\begin{table}[!t]
\begin{center}
\begin{tabular}{lcc}
%\hline
&ESR&CSRe\\
%\hline
Circumference [m]&108.4 &128.8\\
Maximal magnetic rigidity $B\rho_{max}$ [Tm] & 10.0 & 8.40 \\
Electron cooling (ion energy) [MeV/u] & 4-430 & 25-400 \\ 
Stochastic cooling (ion energy) [MeV/u] & 400 & - \\
Transition point $\gamma_t$ (isochronous mode) & 1.41 & 1.40 \\
Transition point $\gamma_t$ (standard mode) & 2.29 & 2.629 \\
Acceptance for cooled ions $\Delta{(m/q)}/(m/q)$  & $\pm$1.5\% & $\pm$1.3\% \\ 
%\hline
\end{tabular}
\end{center}
\caption{Some major parameters of the ESR and CSRe storage rings~\cite{ESR,Xiao}.}
\label{rings}
\end{table}
%---------------------

\section{Nuclear Physics Experiments}
%{\bf All, please, have a look!}\\

\subsection{High-Precision Mass Measurements}
\label{S:masses}

In-ring mass measurements of exotic nuclei were pioneered in the 1990s at GSI~\cite{ims_prop, sms_prop, Geissel92},
and are now conducted at both, ESR and CSRe storage rings.
For ions stored in a storage ring holds the following relationship, which connects the relative revolution frequencies ($f$) or revolution times ($t$) of the circulating ions 
to their relative mass-over-charge ratios and velocities~\cite{FGM,Ra-PRL}:
\begin{equation}
\label{sms1}
\frac{\Delta f}{f}=-\frac{\Delta t}{t}=-\alpha_p\frac{\Delta ({m}/{q})}{{m}/{q}}+(1-\alpha_p\gamma^2)\frac{\Delta v}{v},
\end{equation}
where $\alpha_p$ is the momentum compaction factor, which characterises
the relative variation of the orbit length of stored particles per relative variation of their magnetic rigidity. 
The $\alpha_p=-1/\gamma_t^2$ is nearly constant over the entire revolution frequency acceptance of the storage ring, 
and $\gamma_t$ is the so-called transition point of a ring~\cite{FGM}.

Equation~(\ref{sms1}) is the basic equation for storage ring mass spectrometry.
In order to determine $m/q$ values of the ions, 
one needs to measure their revolution frequencies or alternatively revolution times.
The second term on the right hand side affects the achievable mass resolving power and has to be made as small as possible.
There are two ways to achieve the latter~\cite{FGM}. 
The first one is to reduce $\delta v/v$ by applying beam cooling.
The revolution frequencies are determined from Fourier transformed noise using 
a non-destructive Schottky pick-up installed in the ring aperture, 
on which the ions revolving in the ring induce periodic mirror charges~\cite{Borer}.
This is the basis of the so-called Schottky mass spectrometry (SMS).
The second method is to tune the ring into the isochronous ion-optical setting 
and inject the ions with energies corresponding to $\gamma=\gamma_t$.
This is the basis of the so-called Isochronous mass spectrometry (IMS)~\cite{ims_prop, IMS1, IMS2}.
In this case the velocity spread of the ions is compensated by the lengths of the closed orbits inside the ring
and the revolution frequencies are a direct measure of the mass-over-charge ratios of the ions.
In the IMS, the revolution times can be measured by a dedicated timing detector~\cite{esrtdet, Mei10}.

The IMS does not require beam cooling and is thus ideally suited for mass measurements of the shortest-lived nuclides.
The nuclide with the shortest half-live measured by the IMS is the isomeric state observed in $^{133}$Sb nuclide at an excitation energy of
$E^*=4.56(10)$~MeV~\cite{Sun2,Sun10}.
The half-life of this isomer in neutral atoms is $T_{1/2}^{\rm atom}=16.54(19)$~$\mu$s~\cite{AME12}.
However, due to the fact that bare $^{133}$Sb nuclei were stored in the ESR, 
all decay modes involving bound electrons were disabled~\cite{Li-PLB} and a half-life in the order of 10~ms is expected~\cite{Sun10}. 

%%%%%%%%%%%%%%%%%%%%%%%%%%%%%%%%%%%%%%%%%%%%%%%%%%
\begin{figure*}[t!]
\centering\includegraphics[angle=-0,width=0.8\textwidth]{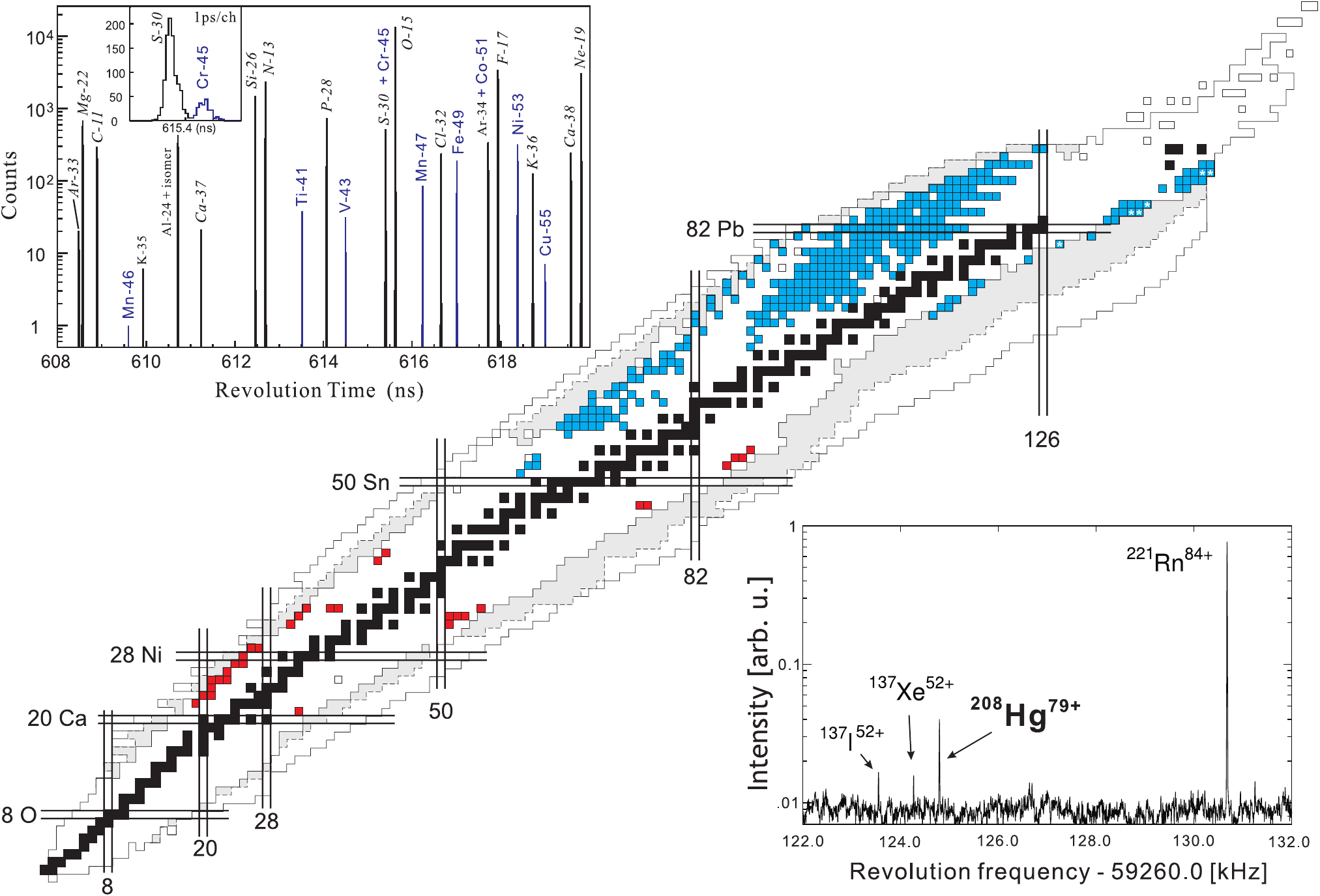}
\caption{%
(Colour online) The chart of the nuclides illustrating nuclides whose masses were measured by storage ring mass spectrometry at ESR and CSRe, 
and those which could be determined by using literature $Q$-values of proton and $\alpha$-decays. 
Only the ground-state masses obtained for the first time are considered.
The white area indicates the nuclides with known masses according to the atomic mass evaluation AME'12~\cite{AME12}.
The grey area with dashed-line borders indicates the range of nuclides which can 
be addressed by storage ring mass spectrometry at the present ESR and CSRe~\cite{YLHX}.  
The outer white region with solid-line borders is the area of nuclides 
that will become accessible at the future FAIR facility; the goal of the ILIMA proposal~\cite{Phil100, FAIRBS, ILIMA}.
The masses stemming from SMS and IMS are indicated by blue and red colours, respectively.
The data on masses addressed by SMS are from~\cite{Ra-NPA, Novikov02, Li-NPA, Chen09, Chen10, Chen12, Shubina13} and
by IMS are from~\cite{IMS2, Zhang12, Yan13, TUNIM, Sun08,  TuPRL, IMS-3}. 
The upper insert shows a part of the revolution time spectrum of $^{58}$Ni projectile fragments measured by IMS in the CSRe. Taken from~\cite{Zhang12}.
The lower insert illustrates a 10~kHz part of a Schottky frequency spectrum of neutron-rich $^{238}$U projectile fragments. 
A single H-like $^{208}$Hg$^{79+}$ ion was seen only once in a two-weeks long experiment. Taken from Ref.~\cite{Chen09}.
}%
\label{nchart}
\end{figure*}
%%%%%%%%%%%%%%%%%%%%%%%%%%%%%%%%%%%%%%%%%%%%%%%%%%
Storage ring mass spectrometry together with cocktail beams offered by the corresponding separators, enables 
broadband investigations of many tens of nuclear species in a single measured spectrum.
For instance a single spectrum in the isochronous mode covers $\Delta (m/q) / (m/q)\approx 13\%$~\cite{IMS1}.
A part of the revolution time spectrum of $^{58}$Ni projectiles addressed by the IMS at CSRe~\cite{Zhang12, Yan13} is illustrated in the upper insert of Fig.~\ref{nchart}.
Many nuclei with previously unknown (roman) and known (italic) masses are simultaneously measured.
The nuclides with known masses provide an {\it in situ} calibration of the revolution time / frequency spectra.

Storage ring mass spectrometry is an extremely sensitive technique.
Often a single stored ion is sufficient to determine its mass with high accuracy.
One example of such measurement is illustrated in Fig.~\ref{nchart}, 
where a Schottky frequency spectrum of stored $^{238}$U projectile fragments is shown in the lower insert~\cite{Chen09}.
The frequency peak at about 125~kHz corresponds to a single $^{208}$Hg nucleus which was seen once as a H-like ion within a two-weeks long experiment.
Furthermore, six new isotopes have been discovered in the ESR together with their mass and half-life measurements.
They are indicated with white asterisks on the chart of the nuclides in Figure~\ref{nchart}.

The sensitivity of the SMS to single stored ions can be used to resolve low-lying isomeric states, which could otherwise not be resolved.
Figure~\ref{125ce} illustrates the discovery of a long-lived isomeric state with an excitation energy of 
$E^*=103(12)$~keV in the neutron-deficient $^{125}$Ce nuclide~\cite{Sun07}.
Combined with the capability of storage rings to cover a wide range of different nuclides in one frequency spectrum, 
the single ion sensitivity can be used for a broadband search of nuclear isomers on the chart of the nuclides.
For instance, a region of neutron-rich nuclei around $^{188}$Hf is predicted 
to exhibit isomers with exceptional properties~\cite{PWalker, PWalkerGD,GD}.
It was recently mapped using SMS~\cite{Shubina13, Reed10, Reed12, Reed12a}.

The harvest of masses obtained for the first time via storage ring mass spectrometry is illustrated on the chart of the nuclides in Fig.~\ref{nchart}.
The achieved relative mass accuracy spans from a few $10^{-7}$ (SMS) to $10^{-6}-5\cdot10^{-7}$ (IMS).
%Only masses of nuclear ground states measured for the first time are considered. The overall covered mass surface includes more than 1000 nuclei.
The new masses enabled numerous investigations of nuclear structure and astrophysics questions which are not possible to cover within this review.
For more details, the reader is referred to Refs.~\cite{BoLiSt, FGM, Ge2001, Bo-IJMS, add2, add3} and references cited therein.

\subsection{Decay Studies of Highly-Charged Radionuclides}
\label{S:lifetimes}

With the FRS-ESR, it became possible to produce unstable, 
highly-charged ions and to inject them into a storage ring~\cite{Geissel92}. 
Here, owing to the ultra-high vacuum, the high atomic-charge states can be preserved and constantly monitored 
for extended periods of time sufficient to study decay properties of such highly-charged ions~\cite{LiBo}.
Interest in such investigations is manyfold:
A straightforward example is that highly-charged ions enable one to study the influence of bound electrons on the radioactive decays.
Here the bare or H-like heavy ions represent 
well-defined quantum-mechanical systems in which the corrections due to otherwise many bound electrons are removed.
Obvious examples are the decays of fully-ionised atoms in which the decays involving electrons are just disabled~\cite{Li-PLB, Irnich}.
Furthermore, new decay modes--strongly suppressed or disabled in neutral atoms--can open up.
Another motivation for such studies are the nucleosynthesis processes in stars, 
where the involved nuclides are usually highly-charged due to high temperatures and high densities of the corresponding environments. 
%For instance in the $s$-process along the valley of $\beta$ stability the mean ``temperature'' (kT) 
%amounts to about 30~keV and in the explosive $r$-process it reaches more than 100~keV.  

We focus here on the studies of weak decays.
Using $p$, $n$, $e^-$, $e^+$ and $\nu_e$ to indicate the proton, neutron, electron, positron and electron neutrino, respectively, 
the latter can be summarised as:
\begin{eqnarray}
p  +  e^-_b   \to    n  +  \nu_e    &   ~~~~{\rm orbital~electron~capture~(EC);}\label{eqec}\\    
p \to n + e^+ + \nu_e & ~~{\rm continuum~\beta^+~decay~(\beta^+_c);}\label{eqbp}\\
n \to p + e^- + \bar{\nu}_e & ~~{\rm continuum~\beta^-~decay~(\beta^-_c);}\label{eqbm}\\
n  +  \nu_e   \to    p +  e^-_b  &      ~~~~{\rm bound-state~\beta-decay~(\beta^-_b).}\label{eqbb}
\label{eq1}
\end{eqnarray}
In the decay the mass-over-charge ratio changes 
which inevitably leads to a change of the revolution frequency in the ring between parent and daughter ions.
In the two-body $\beta$-decays (\ref{eqec}) and (\ref{eqbb}), the charge state is not altered and the frequency change is small and 
directly reflects the decay $Q$-value.
In the three-body decays (\ref{eqbp}) and (\ref{eqbm}) the charge is modified as well which results in a much bigger frequency change. 

\begin{figure}[h]
\centering\includegraphics[angle=-0,width=0.45\textwidth]{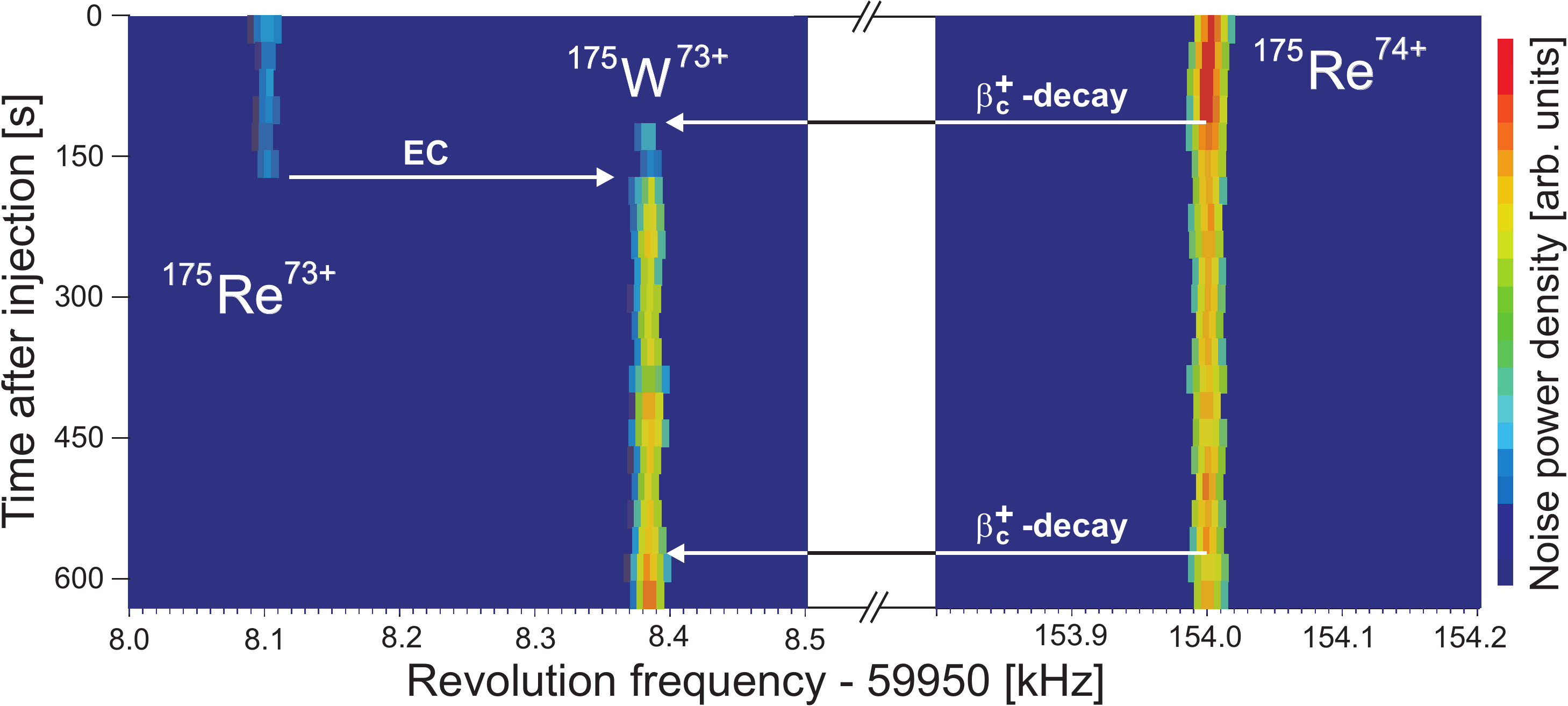}
\caption{%
(Colour online) Time-resolved Schottky frequency spectra of stored isobars with $A = 175$. 
Three H-like $^{175}$W$^{73+}$ ions are created in radioactive decays in the ESR. 
One ion is produced in the EC decay of a single He-like (He-like) $^{175}$Re$^{73+}$ ion at about 3~min after injection. 
The other two ions are created in three-body $\beta_c^+$-decay of two out of a few stored H-like $^{175}$Re$^{74+}$ ions 
at about 2 and 10~min after the injection. 
Taken from~\cite{LiBo}.
\label{figec}
}%
\end{figure}

In case that the frequencies of both, the parent and the daughter ions lie within the storage acceptance of the ring, 
they can be addressed by the time-resolved SMS~\cite{Li04}.
An example is illustrated in Fig.~\ref{figec}, where the daughter H-like $^{175}$W$^{73+}$ ions are populated via 
EC and $\beta^+_c$ decays of respectively He-like $^{175}$Re$^{73+}$ and H-like $^{175}$Re$^{74+}$ parent ions. 
%%%%%%%%%%%%%%%%%%%%%%%%%%%%%%%%%%%%%%%%%%%%%%%%%%
%%%%%%%%%%%%%%%%%%%%%%%%%%%%%%%%%%%%%%%%%%%%%%%%%%

Particle detectors inserted into the ring aperture after a dipole magnet 
can be employed to detect daughter ions if their orbits lie outside the ring acceptance.
In such case, the beam of parent ions, the intensity of which is monitored by the SMS, circulates undisturbed on the central orbit 
of the ring and the created daughter ions are deflected by the magnet and are intercepted by the detector.
In both cases a redundant measurement is achieved in which the decay curve of the parent ions 
and the growth curve of the daughter ions are simultaneously measured.

Several exciting experimental results have been obtained at the ESR.
The bound-state $\beta^-_b$ decay was experimentally discovered in fully-ionised $^{163}$Dy atoms.
Neutral $^{163}$Dy are stable but decay with $T_{1/2}=33$~days if stripped off all bound electrons~\cite{Ju-PRL}.
This measurement led to a determination of the temperature in the $s$-process and set an upper limit for the mass of the electron neutrino.
Another example is the $\beta^-_b$ decay of $^{187}$Re$^{75+}$ nuclei~\cite{Bo-PRL}.
Neutral $^{187}$Re atoms decay with $T_{1/2}=42\cdot10^9$~years to $^{187}$Os atoms.
The decay energy $Q_{\beta^-_c}=2.7$~keV is the smallest known $Q_{\beta^-_c}$ value~\cite{AME12}.
However, the bare $^{187}$Re nuclei decay by merely $T_{1/2}(^{187}{\rm Re}^{75+})=33(2)$~years,
which is by more than 9 orders of magnitude shorter than the half-life of neutral atoms.
Being a pure $r$-process nucleus $^{187}$Re shields $^{187}$Os, which can thus not be produced in the $r$-process.
Therefore, owing to the long half-life of $^{187}$Re, 
it was suggested to use the $^{187}$Re/ $^{187}$Os pair as a cosmic clock  to determine the age of the Universe.
Since $^{187}$Re can be present in different charge states during the galactic evolution, 
the much faster $\beta^-_b$-decay has to be taken into account~\cite{Kohji-AIP}. 

%In recent experiments, $\beta^-_b/\beta^-_c$ ratios could be measured in neutron-rich nuclei in 
%analogy to the $EC/\beta^+_c$ ratios measured on the neutron-deficient side.
In recent experiments, $\beta^-_b/\beta^-_c$ ratios could be measured for the first time in neutron-rich nuclei.
Such ratios are analogous to the $EC/\beta^+_c$ ratios measured on the neutron-deficient side and can be used to test the $\beta$-decay theory~\cite{Bambynek,Faber}.
Decays of bare $^{206}$Tl$^{81+}$, $^{207}$Tl$^{81+}$~\cite{Ohtsubo} and $^{205}$Hg$^{80+}$~\cite{Kurcewicz} have been studied.
Except for a slight deviation in the latter case, the decay rates can well be described by using standard $\beta$-decay calculations.

Measurements of allowed $1^+\to0^+$ Gamow-Teller EC-decay of H- and He-like $^{140}$Pr and $^{142}$Pm ions yielded a striking result~\cite{Li-PRL07, Wi-PLB}.
The EC-decay rates in H-like ions turned out to be about 50\% larger than in the corresponding He-like ions, though
the number of the bound electrons is reduced from two to one.
This counterintuitive effect could be explained if the total angular momentum of nucleus plus lepton system is considered~\cite{Patyk2008a, Ivanov2008e, Patyk2}.
Dependent on the spin-parities of the parent and daughter nuclei as well as on the magnetic moment of the former, 
the allowed EC-decay can be disabled in H-like ions though a bound electron is present~\cite{LiIJMP,add4}.
The first attempt to confirm this was done by studying the decay of highly-charged $^{122}$I ions~\cite{Atanasov12}. 

Another, yet unexplained but broadly discussed in literature, effect of the EC-decay studies of H-like $^{140}$Pr and $^{142}$Pm ions in the ESR 
is the observed 7~s modulation superimposed on the exponential decay curve~\cite{osc, osc13}.

\subsection{In-Ring Nuclear Reactions}
\label{S:reactions}

First in-ring nuclear reaction studies were conducted in the last few years in the ESR.

The first example is the proton capture reaction measurement relevant for the $p$-process of nucleosynthesis.
The favoured sites for the $p$-process are the explosively burning O~/~Ne layers in the type-II Supernovae and the explosive carbon burning in the type-Ia Supernovae,
which last for about 1~s and can be characterised by high temperatures of $2 - 3 \cdot 10^9$~K and densities of about $10^6$~g/cm$^3$~\cite{p-process, p-process2}.
The astrophysical $p$-process involves about 2000 nuclei connected by more than 20000 reactions, mainly $(\gamma, n)$, $(\gamma, p)$ or $(\gamma, \alpha)$.
However, only a handful of experimental data for stable isotopes has been determined in the Gamow window of the $p$-process so far.

Storage rings offer the possibility to address capture reactions on unstable ion beams. 
The proof-of-principle experiment addressing a proton capture reaction relevant for the astrophysical $p$-process was performed in 2009~\cite{Zhong11}.
A primary beam of $^{96}_{44}$Ru projectiles was fully-striped of electrons in an 11~mg/cm$^2$ carbon foil placed in the SIS-ESR transfer line. 
The fully-ionised atoms were injected at 100~MeV/u, stored, electron cooled and decelerated to 9, 10 or 11~MeV/u.
About $5\cdot10^6$ $^{96}$Ru$^{44+}$ ions were stored and cooled at the final energy.
Taking into account the revolution frequency of about 500~kHz and the thickness of the H$_2$ target of about $10^{13}$ atoms/cm$^2$, 
a luminosity of about $2.5\cdot 10^{25}$~/cm$^2\cdot$s is achieved.
The main reaction channel in the target is the atomic electron pick-up (REC) from the target atoms which is accompanied by an emission of an X-ray~\cite{Eichler}.
Moreover, the charge state of $^{96}$Ru ions is changed from $44+$ to $43+$ and the $^{96}$Ru$^{43+}$ ions 
can be detected by particle detectors inside the first dipole magnet downstream of the gas-jet target (see Fig.~\ref{frs_esr}).
Since the K-shell REC cross sections are known to about 5\%, 
the X-rays detected in coincidence with $^{96}$Ru$^{43+}$ ions provide an {\it in situ} measurement of the luminosity.
The nuclear reaction products of $(p,\gamma)$, $(p,n)$ and $(p,\alpha)$ reactions, the $^{97}$Rh$^{45+}$, $^{96}$Rh$^{45+}$ 
and $^{93}$Tc$^{43+}$ ions, respectively, 
are bent to inside orbits by the dipole magnet and are detected by position sensitive double-sided silicon strip detectors (DSSD).
The DSSDs are installed in air in vacuum pockets separated from the ring vacuum by 25~$\mu$m stainless steel windows.
The measured $^{96}$Ru$(p,\gamma)^{95}$Rh cross-section at 10~MeV/u was determined to be $3.6(5)$~mbar.

Unfortunately, the daughter ions with energies of below 10~MeV/u were stopped in the vacuum window and the gas in front of the DSSD, 
which did not allow the measurements directly in the Gamow window of the $p$-process.
Further investigations of $(p,\gamma)$ as well as $(\alpha,\gamma)$ reactions are proposed for the ESR~\cite{E062}.

The second example addresses the physics of X-ray bursters.
Astrophysical X-ray bursts have been interpreted as being generated by 
thermonuclear explosions in the atmosphere of an accreting neutron star in a close binary system~\cite{X-ray}. 
In between the bursts, energy is generated at a constant rate by the hot CNO 
cycle driven by the in-flow of hydrogen and helium material from the less evolved companion star. 
The $^{15}$O$(\alpha,\gamma)^{19}$Ne reaction is a probable candidate for a breakout reaction from the CNO cycle, which then fuels the $rp$-process.
The latter can result in the production of neutron-deficient nuclei, possibly up to Sn-Sb elements~\cite{Schatz}.
The $^{15}$O$(\alpha,\gamma)^{19}$Ne reaction rate controls the conditions for the ignition of the X-ray burst.
This is a resonant reaction, in which the key resonance is at $E_r = 504$~keV, corresponding to a $3/2^+$ state at an excitation energy of 4.033~MeV in $^{19}$Ne.
This state decays predominantly by $\gamma$-decay, and a very weak $\alpha$-decay branch of about $10^{-4}$ was predicted to exist~\cite{Langanke}.
The latter is the key quantity which determines the resonance strength and which has to be verified experimentally.

Now, a proof of principle experiment has been proposed~\cite{Woods} and conducted in the ESR in October 2012 addressing 
the population of the key 4.033~MeV state in $^{19}$Ne via the $(p,d)$ reaction on stable $^{20}$Ne nuclei.
Fully-ionised $^{20}$Ne$^{10+}$ atoms were stored and cooled in the ESR at 50~MeV/u.
A hydrogen gas-jet target with a density of $10^{13}$~atoms/cm$^2$ has been employed.
A DSSD detector was inserted on the inner side of the ESR 50~cm downstream the interaction point.
This detector was used to measure the energies of the emitted deuterons.
A good energy resolution could be achieved online.
The decay products of the $^{19}$Ne$^*$ ions, $^{19}$Ne$^{10+}$ or $^{15}$O$^{8+}$ ions, after $\alpha$-decay, 
were detected with a PIN-diode detector~\cite{PIND} mounted on the outside 
of the ESR about 8~m downstream the interaction point.

In an ideal case, for each detected deuteron corresponding to the populated 4.033~MeV state in $^{19}$Ne$^*$, 
a $^{19}$Ne$^{10+}$ or $^{15}$O$^{8+}$ ion shall be detected in coincidence.
We note that this is the first transfer reaction ever measured in the ESR.
The data analysis is in progress.

Another example concerns the light-ion induced direct reactions.
Such reactions, like for example elastic and inelastic scattering, transfer, 
charge-exchange, or knock-out reactions, have been shown in the past, 
for the case of stable nuclei, to be powerful tools for obtaining nuclear structure information. 
In the last two decades they have also been used for the investigation of exotic nuclei 
with radioactive beams in inverse kinematics. 
In particular, it turned out that in many cases essential 
nuclear structure information is deduced from high-resolution measurements at low momentum transfer. 
For the case of inverse kinematics experiments with radioactive beams, such measurements can be favourably performed with radioactive beams, stored and cooled in storage rings, 
and interacting with thin internal gas-jet targets. 
This technique enables, due to the thin windowless targets and the beam cooling, 
to perform high resolution measurements, even for very slow target-like recoil particles, 
obtained from reactions at low momentum transfer with reasonable luminosity by profiting from the accumulation and recirculation of the radioactive beams~\cite{PEPS, FAIRBS}.

The full potential of this new technique for nuclear structure and nuclear astrophysics studies in the 
region far off stability will only be available once the experimental conditions provided by the future 
FAIR facility with high intensity exotic beams from the new fragment separator Super-FRS (see Section~\ref{S:EXL}) have been realised~\cite{FAIRBS}. 
Nevertheless, pioneering proof-of-principle experiments started recently at the ESR
employing stable $^{58}$Ni and radioactive $^{56}$Ni beams.
% with radioactive beams close to the line of stability produced at the FRS. 
Interactions of these beams with internal 
%present frag-ment separator FRS. The intention of these experiments was twofold: to perform 
%feasi-bility studies with stable 58Ni beams interacting with 
hydrogen and helium gas-jet targets were used to study the experimental conditions for reaction experiments in the environment of a storage ring.
The angular distribution for elastic proton scattering from the doubly-magic $^{56}$Ni was measured 
in order to obtain a deeper insight into the structure of this nucleus, which is of high interest from the nuclear structure and nuclear astrophysics points of view. 
It should be pointed out that this experiment was, even on a world-wide scale, the first of this kind performed with a radioactive beam.
%, and thus represents an essential milestone towards the realisation of the full EXL pro-ject.
%%%%%%%%%%%%%%%%%%%%%%%%%%%%%%%%%%%%%%%%%%%%%%%%%%
\begin{figure}[h]
\centering\includegraphics[angle=-0,width=0.45\textwidth]{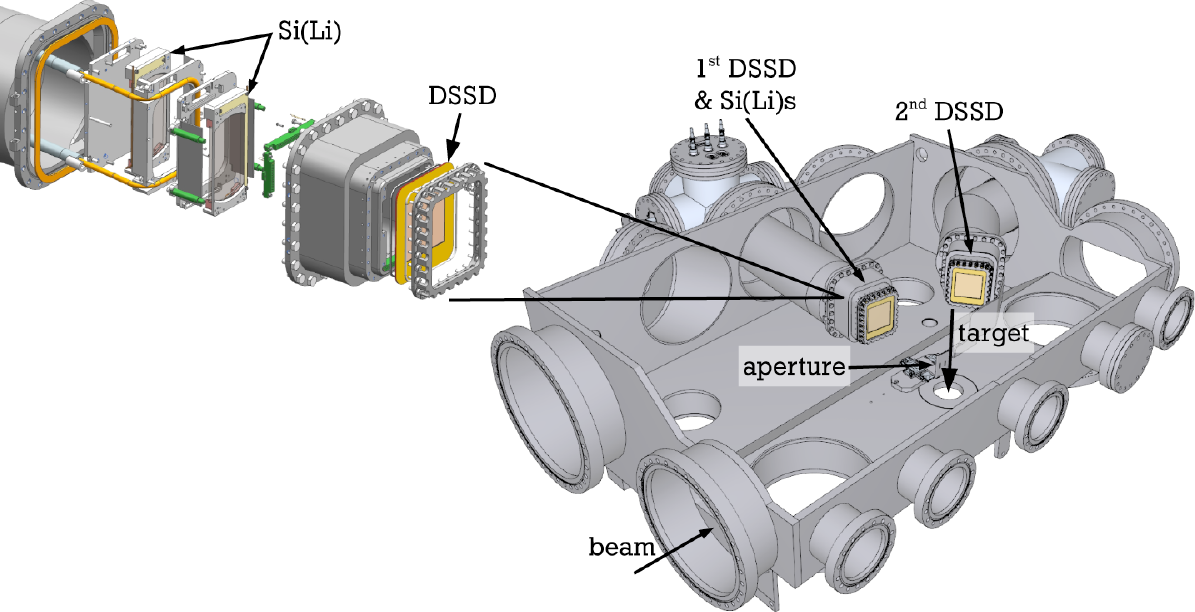}
\caption{%
(Colour online) Schematic view of the EXL setup recently used for reaction experiments at the ESR~\cite{Mirko}. For details see text.
\label{exl_setup}}%
\end{figure}
%%%%%%%%%%%%%%%%%%%%%%%%%%%%%%%%%%%%%%%%%%%%%%%%%%
%%%%%%%%%%%%%%%%%%%%%%%%%%%%%%%%%%%%%%%%%%%%%%%%%%
%\begin{figure}[h]
%\centering\includegraphics[angle=-0,width=0.45\textwidth]{EXL2.pdf}
%\includegraphics[angle=-0,width=0.3\textwidth]{CRYRING.pdf}
%\caption{%
%The energy of the recoil protons versus the laboratory scattering angle as measured by the detector telescope 1 for (p,p) and (p,p`) 
%reactions on 56Ni in inverse kinematics at Elab = 400 MeV/u [M. von Schmid et al.,GSI annual report 2012, to be published].
%}%
%\label{nchart}
%\end{figure}
%%%%%%%%%%%%%%%%%%%%%%%%%%%%%%%%%%%%%%%%%%%%%%%%%%

Within the last years a dedicated innovative experimental setup was designed and 
constructed for these experiments on the basis of the results obtained within the R\&D investigations for the EXL project~\cite{FAIRBS}. 
The setup (see Fig. 1)
%, designed as a prototype version of the recoil detector for the EXL project, 
consists of a UHV-compatible detector chamber which allows for installing, besides the internal gas jet target, 
UHV-compatible detectors for target-like reaction products in a relatively large angular range, 
as well as the necessary infrastructure for performing direct reaction experiments.
At present, two detector units, one consisting of a telescope of a DSSD 
with an active area of $64\times64$~mm$^2$, and two 6.5~mm thick Si(Li) detectors, 
and the other of only one DSSD were installed and used for first measurements.
%, covering the angular range  
%$72^\circ < \Theta_{\rm lab} < 88^\circ$, and $27^\circ < \Theta_{\rm lab} < 38^\circ$, 
%respectively, 
%Here $\Theta_{\rm lab}$ denotes the opening angle between the beam axis and the direction of the ejectile.
It is noted that the DSSDs were used as active windows, 
separating the UHV from an auxiliary vacuum in a vacuum sealed pocket, in which additional Si(Li) detectors, 
as well as all cabling for the readout of DSSDs and Si(Li)s were housed~\cite{bake}. 
%In order to fulfil the strict vacuum conditions in a storage ring a new idea [B. Streicher et al., Nucl. Instr. Meth. A654 (2011) 604] 
%for using Si detector telescopes including their readout and infrastructure in this environment was conceived, tested, 
%and recently successfully applied for the first time under running conditions: DSSDs were used as active windows, 
%separating the UHV from an auxiliary vacuum in a vacuum sealed pocket, in which additional Si(Li) detectors, 
%as well as all cabling for the readout of DSSDs and Si(Li)s were housed. 
In addition, the Si(Li)s were cooled during bake-out of the whole setup, as well as during operation. 
In order to reach the necessary angular resolution required by the experiment, 
%presently limited by the extension of the gas jet target, 
a remotely-controlled moveable aperture
%, mounted on two UHV-compatible piezo motors, 
was placed in front of the target. 
%Besides improving the angular resolution, this system was also very helpful for determining the actual target position. 
%For an online and offline energy calibration and functionality test of the DSSDs, 
%an Am alpha source (not shown in Fig. 1) could be placed in front of the detection systems. 
%In addition, a set of six PIN-diodes (each 10 x10 mm2), 
%moveable as close as possible to the beam for measurements, 
%and outside the beam orbit during beam injection, 
%was positioned in the UHV at a distance of about 8 m 
%downstream from the target for detection of beam-like reaction products in coincidence with the recoil 
%particles [Y. Ke et al., Proceedings of the STORI11 conferenceÉ.].
The beam-like reaction products were detected in coincidence with the recoil 
particles by the PIN diode detector~\cite{PIND} used also in the $^{20}$Ne experiment.

%The present setup was recently mounted at the ESR, 
%commissioned and used for first reaction experiments with stored radioactive beams at an incident energy of E = 400 MeV/u. 
A $^{56}$Ni beam with an intensity of about $7\cdot 10^4$ particles per spill was 
produced by fragmentation of $^{58}$Ni projectiles in the FRS and then injected at an energy of 400~MeV/u into the ESR. 
After stochastic cooling, bunching and stacking~\cite{FNGSI} 
a stored and cooled beam of about $5 \cdot 10^6$ $^{56}$Ni ions was available, which
% in the ESR for experiments. 
resulted--taking into account the density of the H$_2$ target of about $2 \cdot 10^{13}$~atoms/cm$^{2}$--in a luminosity of about $2 \cdot 10^{26}$~/cm$^{2}\cdot$s.
The data analysis is in progress. 
Preliminary results on the elastic and
inelastic proton scattering from $^{56}$Ni can be found in~\cite{Mirko}.
%As an example, preliminary data for elastic and
%inelastic proton scattering from 56Ni at 400 MeV/u [M. von Schmid et al., GSI annual report 2012, to be published] are displayed in Fig. 2. 
%Here the reconstructed energy of the re-coil protons obtained from detector telescope 1 is plotted versus the scattering angle. 
%The kinematical lines for both, elastic and inelastic scattering to the first excited state of 56Ni at 2.7 MeV are clearly visible. 
%From the elastic scattering data, obtained for a rela-tively large angular range covering the region from small angle scattering 
%up to the second minimum, we expect to obtain detailed information on the radial shape of the nu-clear matter distribution of 56Ni. 
%The analysis of a second experiment, performed with stable 58Ni beam interacting at 100 MeV/u with an internal He target, 
%where the intention was to identify very low energy (200-300 keV) 4He recoils from the excitation of the 
%isoscalar giant monopole resonance in 58Ni in DSSD 1 placed at forward angles, is in progress.
%It should be pointed out that, besides the physics interest, the present experiments served 
%also as a proof of principles for the experimental concept of the EXL recoil de-tector, 
%which can be considered as fully successful. For the next series of experiments planned at the ESR and at the CRYRING 
%an upgraded detector setup including 10-20 individual detectors, 
%and thus covering a considerable larger angular range, is presently under consideration.

\subsection{Dielectronic Recombination on Exotic Nuclei}
\label{S:dr}
\begin{figure}
\centering
\includegraphics[width=85mm]{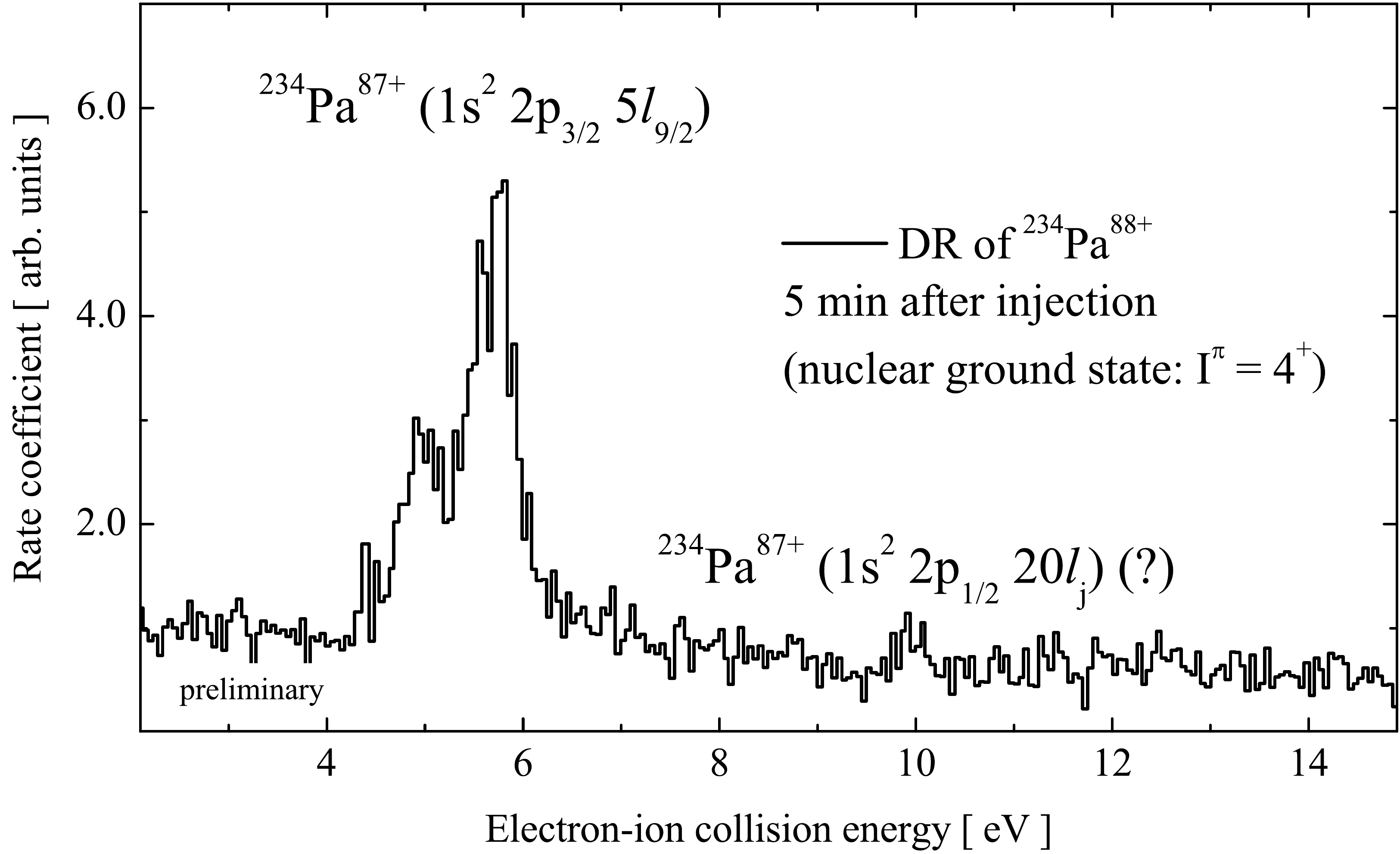}
\caption{
Low-energy DR spectrum of the radioisotope $^{234}$Pa$^{88+}$ taken 5 min after injection~\cite{Brandau13}.  The $^{234}$Pa 
isomers have decayed predominantly by $\beta$-emission. The remaining $^{234}$Pa ions 
are in their nuclear and hyperÞne ground states. The spectrum was obtained with about $2-3 \cdot 10^4$ stored ions. 
The energy calibration is preliminary. Taken from~\cite{Brandau13}.
\label{F:DR}}
\end{figure}
A comparatively new approach to perform isotope shift (IS) and hyperfine splitting experiments of stable and artificially 
synthesised radioisotopes as well as lifetime measurements of long-lived nuclear isomeric states relies 
on the atomic resonant electron capture process of dielectronic recombination (DR)~\cite{Brandau10, CRYRING2, Lestinsky08, Brandau08, Brandau12, Brandau13}.
%[Schuch2005, Lestinsky2008, Brandau2008, Brandau2009, Brandau2010, Brandau2012, Brandau2013].  
%In a nutshell, DR can be regarded as Auger spectroscopy in inverse kinematics since 
%autoionization and the initial dielectronic capture (DC step of DR) are detailed-balance pairs. 
In DR, free electrons with matching kinetic energy are captured resonantly with simultaneous excitation 
of an already bound electron resulting in a doubly excited autoionizing atomic configuration. 
At storage rings, DR is studied with great success, high resolution, precision and sensitivity 
at an electron cooler or a dedicated co-propagating (merged-beams) free-electron target~\cite{Brandau12, Graham02, Muller08}. 
The relative velocity of electrons and the circulating ion beam is tuned in very fine energy steps thus scanning a particular energy range. 
If the resulting relative collision energy matches the resonance condition an enhanced 
number of recombination reaction products, that is, ions with an extra electron, can be registered in the particle detectors in the bending arcs of the storage ring. 
DR is an important atomic collision mechanism that is present in all kinds of plasmas and that is 
of particular importance 
in the highly-ionised matter dominating in astrophysics~\cite{Muller08, Muller12}. 
%The study of DR as a process and the application of its spectroscopic properties belong to the 
%very genuine experiments at heavy ion storage rings [Graham2002]. 
%Over the last two-and-a-half decades a vast progress has been achieved in the understanding of DR,--often in close collaboration of experiment and theory. 
%Based on this sound theoretical background a broad spectrum of topics has been addressed, 
%for instance, in astrophysics, in structure and dynamics of relativistic atomic physics 
%or in fundamental interactions, especially in quantum electrodynamics (QED). 
For a current review about the application of DR as a spectroscopic tool see~\cite{Brandau12}.
%that provides examples, 
%more details about the measurement scheme and further references, cf. [Brandau2012]

\begin{figure}
\centering
\includegraphics[width=85mm]{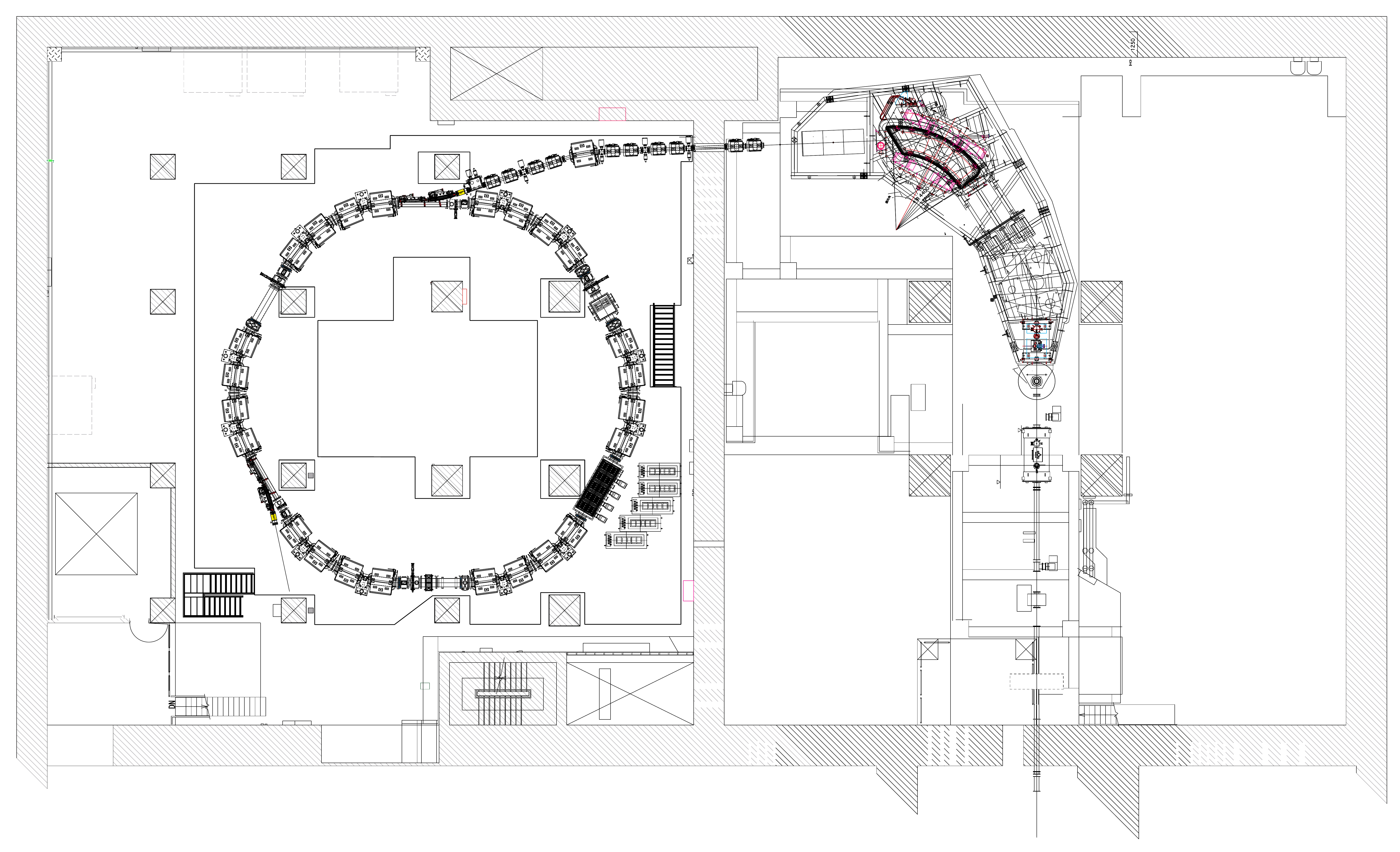}
\includegraphics[width=80mm]{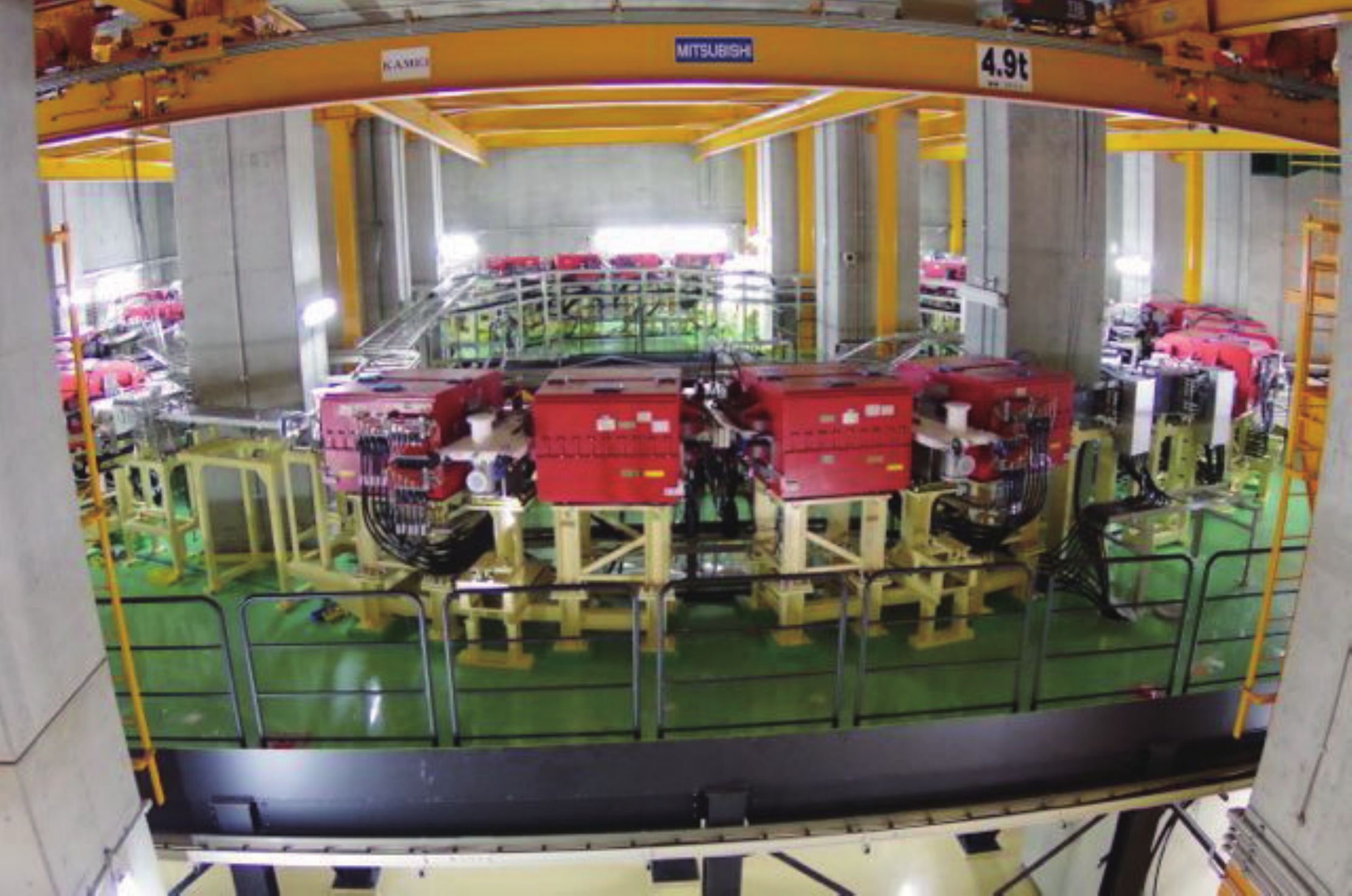}
\caption{
(Colour online) Rare RI-Ring built at the RIBF. The upper panel shows the beam line layout from the SHARAQ spectrometer
to the storage ring, and the lower panel shows a photo of the Rare RI-Ring taken in March, 2013.
\label{R3}}
\end{figure}
In the last decade, the DR has also been applied for the study of 
nuclear properties such as nuclear charge radii, nuclear spins and nuclear magnetic moments. 
The high sensitivity of the DR technique
%storage ring approach 
allows for experiments with as few as $10^2$ to $10^3$ 
stored particles and is thus ideally suited for the investigation of artificially produced radioisotopes. 
In the latter case, a lower limit for the lifetime of the isotopes is given, due to the time needed to store and cool the ions, 
that is, roughly 5 seconds. 
Although at first sight the resolution is considerably worse than in optical and/or laser isotope shift experiments, 
for heavy ions a decisive advantage arises from the free choice of charge-state for the nucleus under investigation. 
In doing so one can tailor the electronic configuration to the needs of the experiment, for DR-IS investigations a 
highly-charged ion with simple atomic structure is usually preferred. 
For heavy ions the many-electrons present for atoms or low-ionisation states 
in laser experiments lead to some ambiguity in the interpretation of the high-quality measurements. 
The present workhorses in DR are ions with Li-like atomic configuration. 
For such three-electron ions the theoretical description of the spectroscopic 
data with respect to nuclear properties can be performed on a full QED level, 
hence allowing for a clear interpretation that is unbiased from atomic many-body effects~\cite{Kozhedub08}.
Early DR-IS experiments were still performed with ions of rather complex electronic structure ($^{207}$Pb$^{53+}$~/~$^{208}$Pb$^{53+}$)~\cite{CRYRING2}, 
but paved the way for the soon-after transition to ions with Li-like configurations. 
Such studies with stable three-electron isotopes were performed at the 
TSR storage ring in Heidelberg on the hyperfine splitting of $^{45}$Sc$^{21+}$~\cite{Lestinsky08} 
and at the ESR at GSI where nuclear volume shift studies of the Nd-isotope couple $^{142}$Nd$^{57+}$~/~$^{150}$Nd$^{57+}$ were performed~\cite{Brandau08}. 

A major breakthrough was achieved at the ESR with the first studies of DR on artificially 
in-flight synthesised radioisotopes such as $^{237}$U$^{89+}$ and $^{234}$Pa$^{88+}$~\cite{Brandau10, Brandau12, Brandau13, Brandau09}.  
The latter isotope is a very interesting case since it possesses a low-energy isomeric level at $73.92+x$~keV with a half-life of 1.17~min 
that decays predominantly by beta-emission~\cite{AME12}. 
This condition renders $^{234}$Pa$^{88+}$ an 
ideal candidate for pilot experiments on DR-IS experiments with nuclear isomers 
since the presence of the isomer can be independently monitored via 
counting of its decay daughters with a particle detector on the inside of the storage ring~\cite{Brandau12, Brandau13}.

Recently, the production of $^{234}$Pa$^{88+}$, the verification 
of its isomeric state in the stored samples and the taking of first 
DR spectra were successful at the ESR (see Fig.~\ref{F:DR}). 
Potential signatures of the isomer were found in the DR spectra and can be verified in a comparison with a 
DR spectrum taken after 5 min when the isomer is fully depopulated (see Fig.~\ref{F:DR}). 
The evaluation of the data is still in progress but already the first online results are very promising~\cite{Brandau13}. 
The monitoring of the DR resonance intensity provides an alternate means
to study the decay of beam composites that cannot be disentangled otherwise~\cite{Brandau12}. 
For example, the method was applied to study atomic metastable states and hyperfine induced decays~\cite{Schmidt94, Schippers07}. 
Similarly, the isomeric decay of long-lived isomers can be addressed even 
with nuclear excitation energies too low to investigate with the majority of other approaches. 
For sure, the most interesting case here is the so-called 'nuclear clock' isomer in $^{229}$Th with 7.8~eV 
nuclear excitation energy connected by an M1 transition to the nuclear ground-state. 
The wealth of $^{229}$Th physics cases and applications is enormous~\cite{EMMI2012}: 
amongst others as a clock with 19 decimal places precision~\cite{Peik03, Campbell12}, 
as an ideal test-bed of fundamental symmetries~\cite{Berengut09}, or as a primary candidate for a nuclear laser~\cite{Tkalya11}. 
Such studies on $^{229}$Th are envisaged at the present GSI installations or at the upcoming new facilities of the FAIR project~\cite{Brandau13}.

\section{Future Storage Ring Facilities}
\label{S:future}

\subsection{Rare-RI Ring Project}
\label{RIRING}
%The r-process is believed to be responsible for the synthesis of approximately half of
%all nuclides heavier than iron.
%However, the pathway is not yet clear, because
%the r-process proceeds in the regions far from the $\beta$-stability line and
%their nuclear properties relevant to the reaction flow are still unknown.
%Among them, nuclear masses are most essential to determine the r-process path.
%Therefore, 
A novel storage-ring project, the ``Rare-RI Ring'', was started at the RIKEN RI Beam
Factory (RIBF), aiming at precision mass measurements of exotic nuclei, in particular, in the vicinity of the $r$-process path~\cite{key2,add5}.
The latter is believed to be responsible for the synthesis of approximately half of all nuclides heavier than iron.

The Rare-RI Ring is a device specially designed for performing isochronous mass measurements on
single exotic nuclei, which have extremely short lifetimes (down to $\sim1$ ms) as well as
extremely low production rates ($\sim1$ count/day).
The radioactive nuclides produced at the fragment separator BigRIPS~\cite{bigrips}
will be identified in-flight on a particle-by-particle basis using position, timing and energy-loss detectors.
The nuclides of interest will then be injected one-by-one into the storage ring by using a specially developed fast-response kicker system.
The ring will be set into the isochronous ion-optical mode with a tuning accuracy in the order of $\delta B/B \simeq10^{-6}$.
Figure~\ref{R3} shows the beam line layout (upper panel) and a photograph (lower panel) of the Rare RI-Ring.
The details of the technique and the presently achieved parameters can be found in a dedicated contribution to this volume~\cite{key1}.

In the Rare-RI Ring IMS measurements,  the mass-over-charge ratio of a stored ion will be determined from 
the revolution time, which will be measured by the fast timing plastic 
scintillation counters located up- and downstream of the storage ring.
A time resolution of better than $\delta T/T \simeq 10^{-6}$ 
(storage time $T\sim1$~ms) is envisioned.
Thanks to the excellent performance of the BigRIPS separator,
a precision velocity measurements can also be performed with an accuracy in the order of $\delta v/v \simeq10^{-4}$.
Combined with the event-by-event velocity correction, the high-precision revolution time
measurements will enable the determination of the masses of short-lived nuclei 
with a relative mass accuracy in the order of $\delta m/m \simeq10^{-6}$.
The experiment is designed such that a single synthesised ion should be sufficient for its accurate mass measurement.
%The IMS measurement in the Rare-RI Ring will in principle be possible on even .
%which allows for studying the nuclei with smallest production rates.

\subsection{Storage ring at HIE-ISOLDE}
\label{TSRISOLDE}
Recently a project has been initiated to couple a 
storage ring to the HIE-ISOLDE facility~\cite{Lin06} at CERN.
An important distinctive feature of this project is that the ion beams will be produced by the Isotope Separation On-Line (ISOL) method, whereas both of the existing storage ring facilities are based on the in-flight production of secondary beams at high energies.
This offers advantages in terms of the beam intensity for a large number of elements and the beam quality.
Moreover, the existing and planned post-acceleration schemes can deliver high-quality ISOL beams right at the required energies, which circumvents the long slowing down times required for the relativistic ion beams.

The Test Storage Ring (TSR)~\cite{Kra90}, which operation was stopped at the Max-Planck Institute for Nuclear Physics 
in Heidelberg~\cite{MPIK} in 2012, is perfectly suited for this purpose.
The physics scope of this instrument includes nuclear physics, nuclear astrophysics and atomic physics, with several experiments that can only be done there.
Examples of the proposed experiments include the measurement of the half-life of $^7$Be in a H-like state, which is of importance for the models of the Sun, 
nuclear structure studies far from stability through nuclear reactions and decay of ions circulating in the ring, research on nuclear isomers, etc.
Also suggested are investigations on nuclear ground-state properties of exotic nuclei via the hyperfine effects on the atomic levels, 
which can be probed with an unprecedented resolution using dielectronic recombination.
The physics motivations are elaborated in detail within an extensive Technical Design Report published recently~\cite{TSR}.
The project has received a skeleton approval by the CERN Research Board, that in August 2012 has also given a one-year mandate to a working group for an integration study of the TSR into the CERN infrastructure.

\subsection{CRYRING at ESR}
\label{CRING}
The Stockholm CRYRING is an immensely successful ion storage ring, 
which has enabled seminal research contributions in atomic and molecular physics for many years. 
As part of the Swedish contribution to FAIR, the CRYRING was transported to GSI.
It is planned to reassemble the CRYRING in the coming two years (see Fig.~\ref{frs_esr}) \cite{CRYRING0,add6}.
With the combination of the ESR and the CRYRING a unique facility will be created, 
which provides cooled, highly-charged ion beams at low energies.  
Thus stored and cooled highly-charged ions up to fully-ionised uranium will be available at GSI in the wide energy range from about 400~MeV/u down to 4~MeV/u in the ESR and then down to a few tens of keV/u in the CRYRING.
Moreover, the CRYRING can operate independently with its own ion source. 
Thus, CRYRING is excellently suited as a test bench for FAIR for
testing technologies and instruments developed for FAIR during the planned shutdown and reconstruction periods of the GSI accelerator infrastructure. 

The CRYRING coupled to the ESR is a powerful scientific instrument 
for research with cooled, highly-charged stable as well as exotic nuclides.
For atomic physics, the low-energy highly-charged beams colliding with electrons and atoms of internal electron or gas targets
will be used as a sensitive spectroscopic tool for the investigation of ionisation, 
recombination, excitation, and resonant scattering.
It will also enable a range of experiments in nuclear and astrophysics as well as at the border between atomic and nuclear physics.
One example of the latter is the search for the long-predicted Nuclear Excitation by Electron Capture process~\cite{Goldanski},
which is a resonant free electron capture into a bound atomic orbital accompanied by the simultaneous excitation of the nucleus. 
This is the exact inverse of the decay of nuclear states by Internal Conversion (IC)~\cite{Palffy}. 
So far, no experimental evidence has been reported for NEEC.

The combination of the ESR and the CRYRING is ideally suited for investigations of astrophysical capture reactions.
The $p$-process Gamow window for capture reactions on nuclei in the tin region at $T_9=2-3$ is $E_{\rm Gamow} = 1.8-4.5$~MeV for proton- 
and $5.3-10.3$~MeV for $\alpha$-induced reactions, which are perfectly within the energy range of the CRYRING~\cite{CRYRINGR,CRYRINGPB}.
%The feasibility of the measurements at $4-10$~MeV/u was demonstrated in the ESR.
These experiments, however, require the installation of particle detectors inside the ultra-high vacuum of the ring.
The development of the corresponding detectors is ongoing. 
Furthermore, reactions of interest for the $rp$-process might be possible to address.
Last but not least, also a wide range of nuclear reaction measurements profiting from 
cooled low-energy radioactive beams is planned in a programme that is complementary to the
studies envisioned by the EXL collaboration at higher beam energies (see Section~\ref{S:EXL}). 

\subsection{Storage Rings at FAIR}
\label{S:FAIR}

\begin{figure}[h]
\centering
\includegraphics[angle=-0,width=0.45\textwidth]{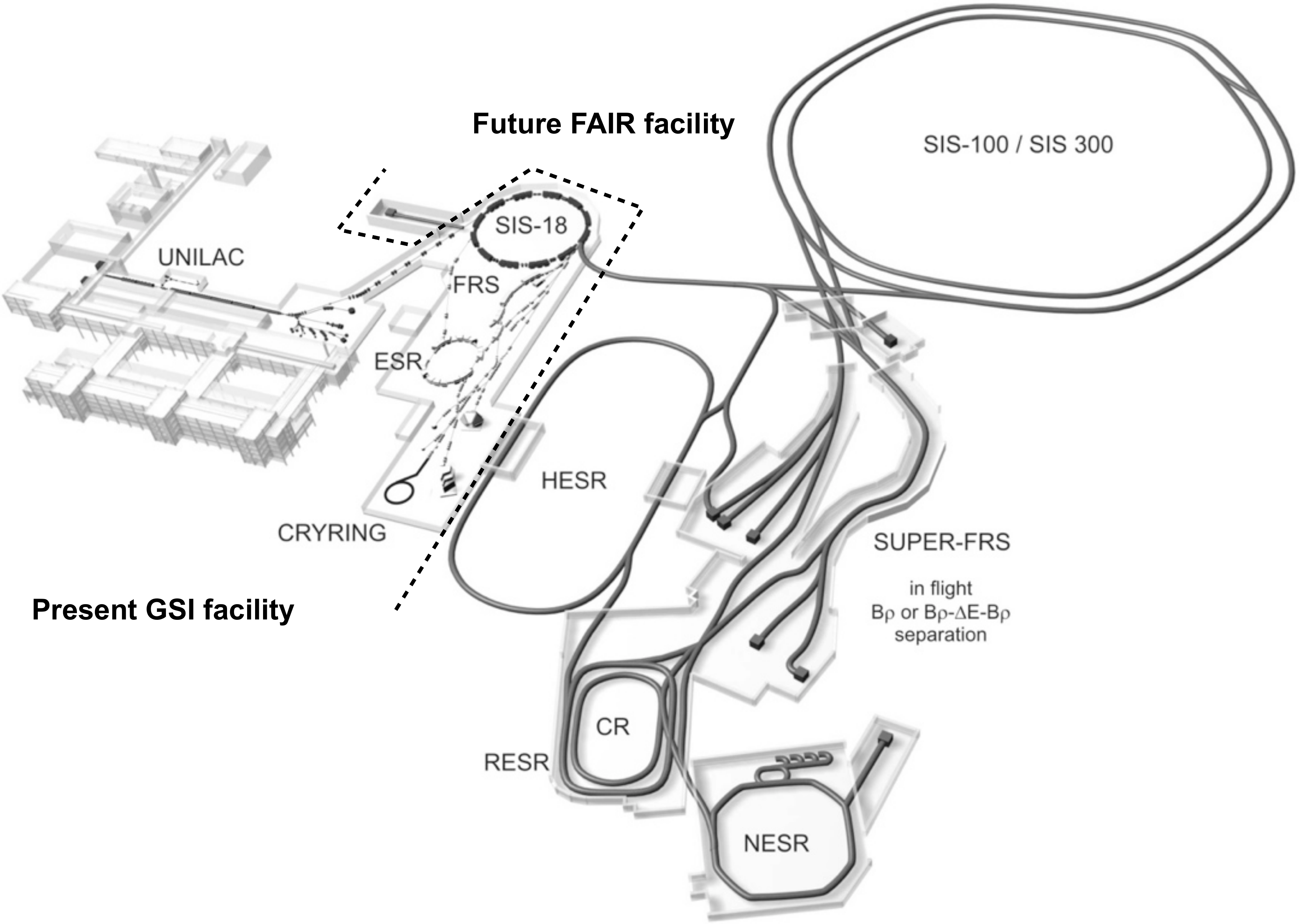}
\caption{%
A schematic view of the Facility for Antiproton and Ion Research in Darmstadt. 
The present GSI facility consisting of the UNILAC, SIS, FRS and ESR is shown together with the location of the CRYRING which is presently being reassembled.
\label{F:FAIR}
}%
\end{figure}

A complex of several storage rings is planned at the future FAIR facility which is schematically illustrated in Figure~\ref{F:FAIR}.
%%%%%%%%%%%%%%%%%%%%%%%%%%%%%%%%%%%%%%%%%%%%%%%%%%
%%%%%%%%%%%%%%%%%%%%%%%%%%%%%%%%%%%%%%%%%%%%%%%%%%

It is proposed to extend the existing GSI facility by adding the 
heavy-ion synchrotrons SIS-100 and SIS-300, a two-stage large-acceptance superconducting fragment separator  Super-FRS~\cite{SFRS} and 
a dedicated complex of storage rings (the Collector Ring (CR), the Recuperated Experimental Storage Ring (RESR),
the New Experimental Storage Ring (NESR), and the High-Energy Storage Ring (HESR))~\cite{SteckIJMP}.

It is envisioned that secondary beam intensities will be superior by about 4 orders of magnitude compared to those presently available.
The exotic nuclei separated in-flight by the Super-FRS will be stochastically pre-cooled in the 
CR and transported via RESR to the NESR or HESR for in-ring experiments.
However, FAIR will be realised in stages, which are defined by the Modularised Start Version of FAIR (MSV)~\cite{MSV}.
The RESR and NESR rings are not part of the MSV and shall be constructed at a significantly later stage.
Due to the MSV, the facility design was modified to enable its operation also without these rings.
%The relevant part of the MSV is illustrated in Figure~\ref{F:MSV_FAIR}.
One of the consequences was that the present ESR will stay in operation until it is replaced by the NESR.
In addition, see Fig.~\ref{frs_esr}, the CRYRING, which was moved from Stockholm University to GSI, 
will be installed behind the ESR~\cite{CRYRING0}.
A beam line connecting the Super-FRS via CR with the ESR is envisaged as an extension of the MSV of FAIR.
If constructed, it will be possible to study the most exotic nuclei provided by the Super-FRS also with detection setups at the ESR-CRYRING.
The experimental conditions at FAIR will substantially improve qualitatively and quantitatively the research potential on the physics of exotic nuclei, 
and will allow for exploring new regions in the chart of the nuclides, 
of high interest for nuclear structure and astrophysics. 
%FAIR will therefore provide world-wide unique opportunities for nuclear structure 
Several scientific programmes are put forward at FAIR and are discussed in the following.

\subsubsection{ILIMA: Isomeric beams, LIfetimes and MAsses}
\label{S:ILIMA}
The ILIMA project is based on the successful mass and half-life measurements at the present ESR.
The key facility here will be the CR, which is particularly designed for conducting IMS measurements~\cite{IsoCR0}.
The ion-optical matching of the Super-FRS and the CR will provide a close to unity transmission of the secondary beams.
The CR will be equipped with two time-of-flight (ToF) detectors installed in one of the straight sections, 
which will enable in-ring velocity measurement of each particle.
The latter is indispensable for correction of the non-isochronicity (see~\cite{Phil100,add7}).
Employing the novel resonant Schottky detectors~\cite{Nolden} will
enable simultaneous broad-band mapping of nuclear masses and lifetimes by the SMS technique.
In addition, heavy-ion detectors will be installed after dipole magnets in the CR. 
The mass surface that will become accessible in the CR is illustrated in Figure~\ref{nchart},
where the smallest production rate of one stored ion per day is assumed.

In addition to the experiments in the CR, there are plans to use the CRYRING and the HESR.
It is proposed to search for the NEEC process in the former, whereas
the accumulation scheme in the latter will be used to achieve high intensities of long-lived highly-charged radionuclides.
One striking example to be addressed is the measurement of the bound-state $\beta^-$-decay of $^{205}$Tl~\cite{205Tl} 
(predicted $T_{1/2}\approx1$~year),
which is important for solar neutrino physics and astrophysics.

\subsubsection{EXL: EXotic nuclei studied in Light-ion induced reactions at the NESR storage ring}
\label{S:EXL}
The objective of the EXL-project, 
is to capitalise on light-ion induced direct reactions in inverse kinematics~\cite{PEPS, FAIRBS}. 
Due to their spin-isospin selectivity, light-ion induced direct reactions at 
intermediate to high energies are an indispensable tool in nuclear structure investigations. 
For many cases of direct reactions the essential nuclear structure information is deduced from high-resolution measurements at low-momentum transfer. 
This is in particular true for example for the investigation of nuclear matter distributions by elastic proton scattering at low $q$, 
for the  investigation of giant monopole resonances by inelastic scattering at low $q$, 
and for the investigation of Gamow-Teller transitions by charge exchange reactions at low $q$. 
Because of the conditions of inverse kinematics in case of beams of unstable nuclei, 
low-momentum transfer measurements turn out to be an exclusive domain of storage ring experiments. 
Here luminosities are superior by orders of magnitude compared to experiments with external targets. 
%%%%%%%%%%%%%%%%%%%%%%%%%%%%%%%%%%%%%%%%%%%%%%%%%%
\begin{figure}[h]
\centering
\includegraphics[angle=-0,width=0.45\textwidth]{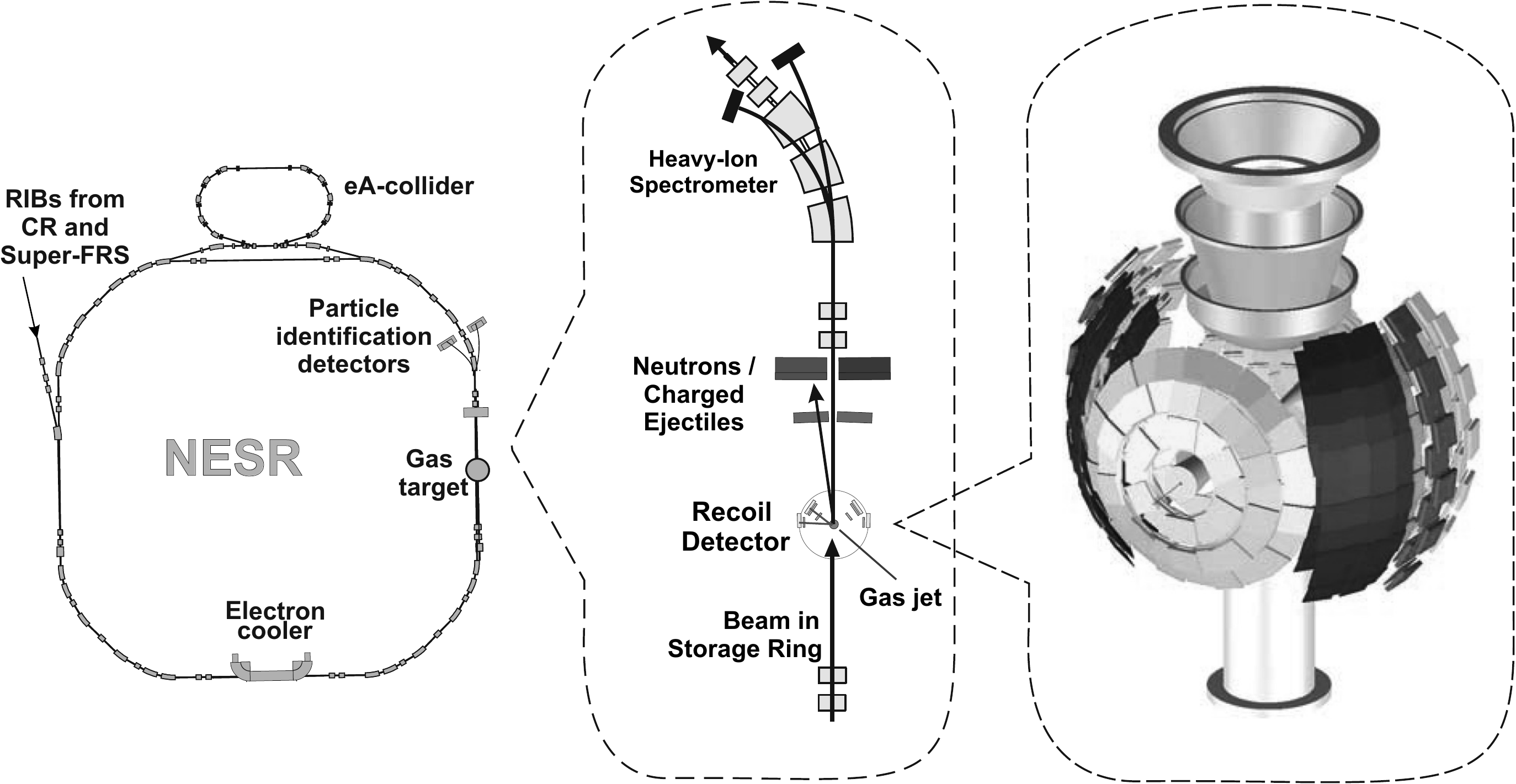}
\caption{%
Layout of the EXL setup at the NESR storage ring. 
The left panel displays a view of the target recoil detector surrounding the internal gas-jet target (for details see text).\label{F:exl3}
}%
\end{figure}
%%%%%%%%%%%%%%%%%%%%%%%%%%%%%%%%%%%%%%%%%%%%%%%%%%

Within the EXL project the design of a complex detector setup was investigated 
with the aim to provide a highly efficient, high-resolution universal detection system, applicable to a wide class of nuclear reactions. 
It is schematically shown in Fig.~\ref{F:exl3}.
The apparatus is foreseen to be installed at the internal gas-jet target of the NESR storage cooler ring.
Since a fully-exclusive measurement is envisaged, the detection system includes: 
a recoil detector consisting of a Si-strip and Si(Li) detector array for recoiling target-like reaction products, 
completed by a scintillator array of high granularity for $\gamma$-rays and for the total-energy measurement 
of more energetic target recoils; a detector in forward direction for fast ejectiles from the excited projectiles, 
i.e., for neutrons and light charged particles; and heavy-ion detectors for the detection of beam-like reaction products. 
All detector components will practically cover the full relevant solid angle and have detection efficiency close to unity. 
One of the most challenging topics will be the design and construction of the recoil detector for target like light ions (see Fig.~\ref{F:exl3}), 
consisting of arrays of Si-strip and Si(Li) detectors, which need to fulfil  demanding experimental conditions with respect to resolution, 
dynamic range, UHV vacuum conditions of storage rings, etc. 
For this purpose an intense R\&D programme was already 
performed~\cite{PIND, bake, EXLNIM, Nasser1, Mirko2}.

Since the NESR is not a part of the MSV of FAIR, the 
realisation of parts of the EXL project in the present ESR is discussed.
Therefore, a transfer line from the CR to the ESR 
needs to be constructed such that the beams from the Super-FRS, pre-cooled in the CR, can become
available also in the ESR.

\subsubsection{ELISe: ELectron Ion Scattering in a Storage ring e-A collider}
\label{S:ELISE}
Electron scattering off nuclei provides a powerful tool for examining nuclear structure, 
which allows precision studies of nuclear properties like charge radii, density distributions, spectroscopic factors as well as giant resonances, etc.
However, the electron scattering experiments are limited today to stable nuclear targets~\cite{ELST}.

The ELISe project aims at elastic, inelastic and quasifree electron scattering, 
which will be possible for the first time for short lived nuclei~\cite{Tom, ELISE}. 
It is proposed to construct an electron-ion collider at the NESR by using the intersecting ion and electron storage rings.
For this purpose a small Electron and Antiproton Ring (EAR), $eA$ collider in Figure~\ref{F:exl3}, 
will be build which will store an intense electron beam (or an antiproton beam, see Section~\ref{S:AIC}) .
The scattered electrons will be analysed with a dedicated electron spectrometer 
with high resolution and large solid angle coverage,
while the recoils will be analysed with particle detectors using the bending section of the NESR as a spectrometer~\cite{ELISE, ELISE2, ELISE3}.
The investigation of charge densities, transition densities, single particle structure, 
etc. on nuclei far off stability will complement the investigations of the EXL project, 
and allow in many cases for a more complete and model independent information of the structure of such nuclei.

Similar to the EXL project, due to the delay in the construction of NESR, 
the feasibility to realise the ELISe project at the present ESR is being investigated.
In this case, the ELISe project will also profit from the CR-ESR transfer 
line delivering secondary beams from the Super-FRS to the ESR.

\subsubsection{AIC: The Antiproton-Ion-Collider at FAIR}
\label{S:AIC}
One of the main capabilities of FAIR is the availability of intense antiproton beams in a wide range of energies.
Within the AIC project, it is proposed to explore collisions of antiproton beams with beams of exotic nuclei,
in order to measure simultaneously neutron and proton distributions~\cite{AIC0, AIC1, AIC2}.
Here, a modified EAR ring (see Section~\ref{S:ELISE}) can be used to store 30~MeV antiprotons.
The antiprotons are brought into head-on collisions with relativistic exotic nuclei stored in the NESR at energies of up to 740~MeV/u.
An antiproton can annihilate on a proton or a neutron of the studied nucleus.
This leads to different reaction-product nuclei which can unambiguously be identified 
after the reaction with Schottky spectrometry or particle detectors after the bending section of the NESR.
Thus, an independent measurement of proton and neutron radii within the same experiment is possible.

The AIC project requires both, the NESR and EAR rings as well as the possibility to bring antiproton beams into the EAR. 
This complicates the transfer of the AIC project to the ESR for the time until the construction of NESR is completed, though various possibilities are presently being studied
taking into account that the transport of antiprotons would be possible with the CR-ESR transfer line.

\subsubsection{SPARC: The Stored Particle Atomic Research Collaboration at FAIR}
\label{S:SPARC}
The SPARC collaboration aims to exploit the multitude of new and challenging opportunities for atomic physics research at FAIR~\cite{SPARC}.
Relevant for this review, is the planned application of the atomic physics techniques to determine properties of stable and unstable nuclei.
Thus, the measurements of isotopic shifts by employing laser spectroscopy or dielectronic recombination studies (see Section~\ref{S:dr}) 
will allow the determination of nuclear magnetic moments, spins and radii.
Furthermore, it was recently proposed to address these properties in a complementary way 
by measuring the degree of linear polarisation of emitted X-rays in highly-charged ions~\cite{Andrey}.

The SPARC research programme was focused on the NESR, whose design was correspondingly optimised.
The development of the required instrumentation and a significant part of the envisioned experiments are feasible in the present ESR, 
especially after the CRYRING will be constructed behind the ESR~\cite{CRYRING0}, and at the HESR~\cite{SPARC_HESR}.
In the latter case, due to the Doppler effect the laser spectroscopy studies will profit from the high kinetic energies by addressing transitions that are inaccessible otherwise.

\subsection{High Intensity Heavy Ion Accelerator Facility}
\label{S:HIAF}

The concept for a new-generation heavy-ion accelerator facility in China, 
High Intensity Heavy Ion Accelerator Facility (HIAF), is being prepared at the IMP. 

For the driver machine of this national user facility it is proposed to use 
a powerful superconducting linac, which shall be able to accelerate intense beams of all elements from protons to Uranium.
The design of the HIAF is not finalised yet.
However, it is envisioned to have a complex of several storage rings which are considered for a broad multidisciplinary research, like
nuclear structure, astrophysics, atomic-, plasma-, accelerator-, and neutrino- physics, as well as investigations of fundamental symmetries and interactions, etc. 
It is emphasised, that a dedicated storage ring for isochronous mass measurements of shortest lived nuclides is envisioned.

\section{Conclusion}
The commissioning of the ESR in 1990 and the first exciting results at this facility resulted 
in a number of storage ring projects which were proposed worldwide.
Examples of such ion-storage rings are for instance the HISTRAP project at ORNL in Oak Ridge~\cite{HISTREP},
the K4-K10 storage ring complex at JINR in Dubna~\cite{K4K10}, a 2-Tm storage ring based on an ISOL-type facility at LBL in Berkeley~\cite{LBL},
and the MUSES storage ring complex at RIKEN in Saitama~\cite{MUSES}.
However, only the storage ring facility CSRe in Lanzhou was taken into operation in 2007~\cite{Xiao}.

In the last two decades, ion storage-cooler rings, fed by highly-charged radioactive nuclei, 
have been proven as excellent tools to study the ground state properties, such as masses and $\beta$-lifetimes, of these exotic nuclei.
The dreams to conduct nuclear reactions in a storage ring were around ever since the ESR was taken into operation~\cite{PEAP, Henning97, Bertulani97}.
The advantages of such studies are the inverse reaction kinematics, the brilliant quality of the cooled beams
and the unique possibility to re-use many times the rare nuclear species for reactions 
with windowless thin internal targets.
Now, the first reaction experiments were successfully conducted in the ESR.

This all led to an increased interest in in-ring experiments addressing nuclear structure and nuclear astrophysics questions,
which is indicated by a number of new storage ring projects that were launched worldwide.
These projects comprise facilities for storing beams at energies of 10~MeV/u and below (the storage ring project at HIE-ISOLDE and CRYRING at the ESR),
at intermediate energies of a few hundreds of MeV/u (the CR, the RESR, the HESR and the RI-RING) 
and at high energies reaching 5~GeV/u and higher (the HESR and the high energy rings within the HIAF project in China).
The highly-complementary physics programmes envisioned at these rings aim at exploiting the unique 
capabilities of the corresponding radioactive beam facilities and will undoubtedly provide new breathtaking results already in the near future.

\section*{Acknowledgements}
This review is entirely based on the common effort of our colleagues working 
with us over many years on various storage ring experiments. 
We appreciate very much their essential contributions and help concerning practical issues as well as enlightening 
discussions about physics which we enjoy.
This research was supported by the DFG cluster of 
excellence ``Origin and Structure of the Universe'' of the Technische Universit{\"a}t M{\"u}nchen,
by the BMBF grant in the framework of the
Internationale Zusammenarbeit in Bildung und Forschung Projekt-Nr. 01DO12012 
as well as by BMBF grants (Contracts 06GI911I and 06GI7127/05P12R6FAN),
by the Helmholtz-CAS Joint Research Group HCJRG-108,
by the Helmholtz Association via the Young Investigators Project VH-NG 627,
by the Max-Planck Society, 
by the Helmholtz Association through the Nuclear Astrophysics Virtual Institute (VH-VI-417), 
by the EuroGenesis activity of the EU,
by the National Natural Science Foundation of China (grant number 11205205),
by the Japanese Ministry of Education, Science,
Sport and Culture by Grant-In-Aid for Science Research under Programme No. A 19204023,
by the the Alliance Programme of the Helmholtz Association (HA216/EMMI),
by HIC-for-FAIR through HGS-HIRE,
and by STFC(UK).



%% References








% \end{thebibliography}

\end{document}